\documentclass[useAMS,usenatbib]{mn2e}
\usepackage[utf8]{inputenc}
\usepackage[T1]{fontenc}
\usepackage{epsfig}
\usepackage{graphicx}
\usepackage{graphics}
\usepackage{color}
\usepackage{amssymb}
\usepackage{amsmath}

\usepackage{caption}

\newcommand{\aap}{Astronomy and Astrophysics}
\newcommand{\mnras}{Monthly Notices of the Royal Astronomical Society}
\newcommand{\apj}{The Astrophysical Journal}
\newcommand{\apjs}{Astrophysical Journal Supplements}
\newcommand{\apjl}{Astrophysical Journal Letters}
\newcommand{\nat}{Nature}
\newcommand{\araa}{Annual Review of Astronomy and Astrophysics}
\newcommand{\apss}{Astrophysics \& Space Science}

\title[EUV sources in old stellar populations]{Emission line diagnostics to constrain high temperature populations in early-type galaxies}
\author[T.E. Woods and M. Gilfanov]{T.E. Woods$^{1}$\thanks{E-mail:
tewoods@mpa-garching.mpg.de} and M. Gilfanov$^{1,2}$\\ $^{1}$Max Planck Institute for Astrophysics, Karl-Schwarzschild-Str.
1, Garching b. M{\" u}nchen 85741, Germany\\
$^{2}$Space Research Institute of Russian Academy of Sciences, Profsoyuznaya 84/32,117997 Moscow, Russia\\}

\begin{document}

\pagerange{\pageref{firstpage}--\pageref{lastpage}} \pubyear{2013}

\maketitle

\label{firstpage}

\begin{abstract}

Once thought to be devoid of warm and cold interstellar matter, elliptical galaxies are now commonly observed to host extended regions of neutral and ionized gas.
Outside of the innermost nuclear regions of these galaxies, the favoured candidate ionizing source remains some component of the stellar population, with mounting evidence suggesting post-asymptotic-giant-branch stars (pAGBs).
In a recent paper, we demonstrated that observations of recombination lines of He II (or upper limits thereof) may provide a strong constraint on the presence of any other, higher temperature ionizing sources, in particular nuclear-burning white dwarfs in the context of the single degenerate (SD) scenario for type Ia supernovae. The sensitivity of the HeII test is greatest for WD effective temperatures $\sim$ $2\cdot 10^{5}K$. 
Here we extend our analysis to include predictions for all of the ``classical'' strong optical lines, as well as UV, optical, and infra-red lines of neutral Oxygen, Nitrogen, and singly-ionized Carbon. This allows us to extend the temperature range over which we can meaningfully constrain the collective luminosity of nuclear-burning WDs to $10^{5}K$ $\lesssim$ T $\lesssim$ $10^{6}K$.
We then demonstrate how observations of nearby early-type and post-starburst galaxies can place strong limits on the origin of type Ia supernovae.

\end{abstract}


\section{Introduction}

It is now well-established that early-type galaxies possess substantial ISM, with neutral hydrogen masses on the order of $\sim$ $10^{6}$ -- $10^{9}M_{\odot}$ \citep{Serra12}.
Accompanying this neutral gas are emission-line regions extending out to several kiloparsecs, with line ratios characteristic of gas in a relatively low state of ionization.
They appear to be powered by the diffuse galactic background, rather than being a superposition of individual H II regions around bright compact sources.
In the past, much effort has been devoted to understanding the sources of ionizing radiation powering these nebulae.
In passively-evolving (non-star-forming) galaxies, it is the old stellar population itself which appears likely to provide the dominant contribution to the local extreme-UV background \citep[e.g][]{Yan12}, at least outside of their innermost nuclei. There, low-luminosity AGN had until recently been thought to provide at least a significant contribution, at least in many cases \citep[e.g. ][]{Annibali10}. However, increasing evidence suggests that the old stellar population may still provide a significant, or even dominant role in powering LINER emission \citep[e.g.][]{CALIFA12}. Within the stellar population, post-asymptotic giant branch stars (pAGBs) remain the favoured candidate for powering the observed emission lines. In particular, pAGB stars are expected to provide sufficient ionizing photons in order to account for the observed H$\beta$ flux \citep{SAURON10}, and the radial surface brightness profiles of the observed emission lines are consistent with ionization by a diffuse population, rather than a central source \citep[e.g.][]{Singh13, CALIFA13}.

At the same time, similar arguments have proven capable of ruling out other possible ionizing sources. 
Recently, \cite{Woods13} proposed that understanding the mechanism powering the diffuse line-emission in passively-evolving galaxies may shed light on the origin of type Ia supernovae (SNe Ia).
There exist two families of models for how these tremendous explosions come about:  in the double-degenerate scenario \citep{Webbink84}, these events result from the merger of a binary pair of white dwarfs (WDs).
In the standard model of the single-degenerate channel \citep{Whelan73}, a carbon-oxygen WD accretes matter from a main sequence or red giant companion until triggering a thermonuclear explosion.
If the latter is correct, then accreting, nuclear-burning WDs should be an extremely luminous component of any stellar population. With temperatures on the order of $10^{5}$ -- $10^{6}K$, they should provide the dominant ionizing background in relatively young ($\lesssim$ 4Gyr) passively-evolving stellar populations.
If this is the case, it should be readily apparent from spectroscopic observations of those ellipticals which host low-ionization emission-line regions \citep{Woods13}. 
This provides a unique opportunity to constrain plausible progenitor models for SNe Ia.

In particular, the observed line flux in recombination lines of ionized Helium can provide an excellent test for the presence of relatively high temperature sources ($\approx$ 2 -- 6 $\cdot$ $10^{5}K$), and their total ionizing luminosity. 
This covers much of the temperature range we expect from 
accreting, nuclear-burning white dwarfs, such as supersoft sources (SSSs), or in particular, the so-called `ultra-soft sources' \citep[those accreting white dwarfs with greatly inflated photospheres, see][]{HKN99}. However, we expect higher temperatures (up to $\sim$ $10^{6}K$) for the most massive ($\gtrsim$ 1.0 $M_{\odot}$) nuclear-burning white dwarfs.
In the past, any possible population of such objects has been severely constrained based on the observed paucity of soft X-ray emission in early type galaxies, and the lack of individual soft X-ray sources \citep[e.g.][]{GB10, DiStefano10}. 
However, independent diagnostic tools remain extremely important.  

Here we extend the work of \cite{Woods13} to include a number of emission-line diagnostic tests for the presence of a population of very hot sources ($T_{\rm{eff}}$ $\gtrsim$ $5\cdot 10^{5}K$), as well as those too cool to strongly ionize He II ($T_{\rm{eff}}$ $\approx$ $10^5K$). In $\S$2, we review our model assumptions for the ISM in early-type galaxies, as well as the possible sources of ionizing radiation in passively-evolving stellar populations.
In $\S$3, we discuss those lines which are most sensitive to ionization by high temperature sources, in particular those of [O I], [N I], and [C II]. 
Together with observations of recombination lines of He II, this extends the range over which emission lines can meaningfully constrain the hardness of the ionizing background in passively-evolving galaxies to cover all temperatures expected from steadily nuclear-burning white dwarfs, including those with inflated photospheres  ($1.5\cdot 10^{5}K$ $\leq$ $T_{\rm{eff}}$ $\leq$ $10^{6}K$). 
In $\S$4, we outline the even more pronounced effect of any putative SD progenitor population (with $\rm{T}_{\rm{eff}}$ $\gtrsim$ $10^{5}K$) in young (t $\lesssim$ 1Gyr), post-starburst populations (such as E+A galaxies). In this case the ionizing luminosity provided by SD progenitors would exceed that available from any other stellar source by up to two orders of magnitude, indicating the specific luminosity in any of the classical strong optical lines (H$\alpha$, [O II] 3727\AA, etc.) should be similarly enhanced. We follow this in $\S$ 5 with estimates of the observational prospects for producing limits using a number of instruments and surveys, as well as how the presence of a significant population of accreting WDs may compromise measures of the SFR in galaxies. 
Finally, we discuss implications for the evolution of binary stellar populations and the progenitors of type Ia supernovae.

\section{Modelling low-ionization emission line regions}

\subsection{Density, metallicity, and ionization parameter}

We model the emission-line regions in early-type galaxies using the 1-D photoionization code MAPPINGS III \citep[see e.g.][]{Groves04}. 
Our procedure follows closely that of \cite{Woods13}; here for convenience we summarize some of the essential ingredients. 

 Unless otherwise stated, we assume a solar \citep{AG89} metallicity for the gas phase, and 2.5 solar for the stellar population. The former is consistent with the mean Oxygen abundance measured in the warm ISM of nearby retired galaxies \citep{AB09, Annibali10}. 
For the ionizing continuum from the `normally-evolving' (single star) stellar population, we use the  population synthesis calculations of \cite{BC03}. 

The most common morphology of the cold ISM in such galaxies is a disky distribution extending from the nucleus out to several tens of kpc \citep{Serra12}, with emission-line nebulae typically ionized by the diffuse background of the stellar population rather than nearby individual sources.
 Therefore, we assume plane parallel geometry throughout. 
For our standard case, we assume a constant hydrogen number density of 100$\rm{cm}^{-3}$, consistent with the observed [S II] 6717\AA/6731\AA\ ratio observed in stacked passively-evolving galaxies from the SDSS \citep{Yan12}, although this may only be an approximate upper bound \citep{Osterbrock}. In principle, the majority of the optical emission lines considered in this work are relatively insensitive to variations in the density.

In our photoionization calculations, we assume that the modeled nebulae are ionization-bounded \citep[see][and references therein]{Woods13}, and terminate our calculations when the ionized Hydrogen fraction falls below 1\%.
The ionization parameter (U = $\dot N_{\rm{ph}}/4\pi r^{2} c n_{\rm{H}}$) is similarly well constrained by the observed ratio of [O III] 5007\AA/H${\beta}$, generally lying in the range -4 $\lesssim$ log(U) $\lesssim$ -3.5 \citep[e.g.][]{Binette94, Yan12}. 

Where appropriate, we normalize the line flux per unit stellar mass of the host population, and assume a covering fraction of unity. This can of course be scaled as needed, since the line luminosity is directly proportional to the stellar ionizing luminosity absorbed by the ISM in ellipticals. In practice, it is often more convenient to provide line predictions normalized to the nearest recombination line of Hydrogen, in order to minimize the importance of reddening in comparison with observed values. This has the added benefit of removing any dependence on the covering fraction of the gas, therefore we also provide predictions for each optical line luminosity relative to the nearest Hydrogen recombination line.

\begin{figure}
\begin{center}
\includegraphics[height=0.25\textheight]{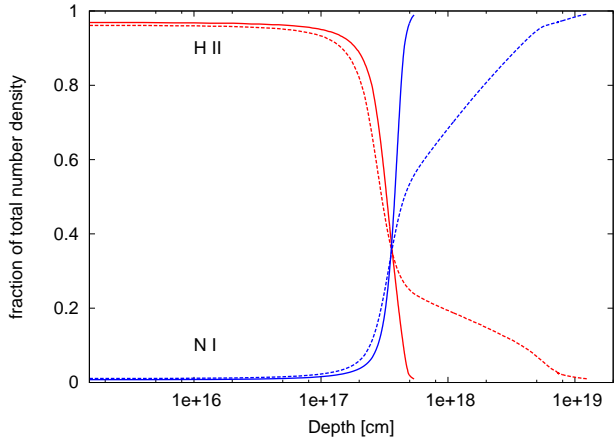}
\caption{Ionization structure of a model nebula, given photoionization by a 3 Gyr-old stellar population (solid lines), and with the addition of our standard example SD SN Ia progenitor population (dashed lines). Red lines indicate the ionized hydrogen fraction, while blue lines follow the neutral fraction of the total Oxygen number density. We assume in both cases log(U) = -3.5. Note that we stop our model calculations when the Hydrogen ionization fraction reaches 1\%. \label{HI_II}}
\end{center}
\end{figure}

\subsection{Available ionizing sources}

In early-type galaxies, pAGBs remain favoured to be the predominant ionizing source (at least within the stellar population). 
Empirically, this has been strongly supported by energetic considerations \citep[e.g.][]{SAURON10}, as well as more detailed emission-line studies \citep{Binette94}. 
The spectral emission from these sources has been studied in detail by \cite{Rauch03}, whose models are included in the spectral synthesis calculations of \cite{BC03} used here. Despite mounting evidence \citep[e.g.][]{Binette94, SAURON10, Yan12}, some doubt remains about the longevity of post-AGB stars, and therefore their expected numbers and contribution to the ionizing background \citep{Brown08}. Therefore, where appropriate, we will consider in our calculations the ionizing background with and without the contribution normally expected from pAGBs. 

Non-stellar sources, such as low-luminosity AGN, likely also play a role in many galaxies \citep{Annibali10}, at least in their central regions. 
Fast shocks have been thought to play a role in some cases, as well as ionization by the hot, X-ray emitting diffuse ISM \citep{SAURON10}. 
However, such contaminating cases can be excluded with the careful use of line diagnostics \citep[e.g][]{Fernandes11}, allowing one to confine their attention to so-called 'retired' galaxies where the old stellar population is the principle source of ionizing emission.

Other stellar sources, such as low-mass X-ray binaries, have been suggested, but do not appear to be significant in this regard \citep{SAURON10}. 
Accreting, nuclear-burning WDs have, however, been found to likely provide a significant contribution, and would be expected to dominate the ionizing contribution for relatively young, passively-evolving stellar populations, at least if they exist in sufficient numbers to account for the observed SN Ia rate \citep{Woods13}. 
Even if they are not the primary evolutionary channel for SNe Ia, accreting WDs may still account for a substantial contribution to the ionizing background (see Chen et al. in prep). 

 The number of nuclear-burning WDs expected in any stellar population, assuming the SD channel accounts for all SNe Ia, can be estimated from the SN Ia rate, the mass accretion rate, and the characteristic mass accreted per SN Ia \citep[as in][]{GB10, DiStefano10}:

\begin{equation}
N_{\rm{SD}} = \frac{\Delta m_{\rm{Ia}}}{\dot M}\dot N_{\rm{Ia}}
\end{equation}

\noindent where $\dot N_{\rm{SN Ia}}$ is SN Ia rate of the galaxy, and $\Delta m_{\rm{Ia}}$ is the total mass accreted prior to explosion ($\approx 0.3$ -- $0.7M_{\odot}$). This depends sensitively on our assumptions regarding the typical accretion rate, which in turn depends on as-yet poorly understood details of binary evolution. Fortunately, we can remove any dependence on the accretion rate if we consider instead the total bolometric luminosity of all accreting, nuclear-burning WDs, $L_{\rm{tot, SNIa}} = L_{\rm{WD}}\cdot N_{\rm{SD}}$. The luminosity of any individual source is simply $L_{\rm{WD}} =  \epsilon \chi \dot M$, with $\epsilon = 6\cdot 10^{18}$erg/s the energy release from nuclear-burning of hydrogen, and $\chi = 0.72$ the hydrogen fraction of solar metallicity gas. Therefore, the total luminosity of any putative SD progenitor population can easily be estimated as:

\begin{equation}
L_{\rm{tot,SD}} = \epsilon _{\rm{H}}\chi \Delta m_{\rm{Ia}} \dot N_{\rm{SN Ia}}
\label{eq:ltot}
\end{equation}

\noindent independent of the assumed accretion rate \citep{GB10}.
For the SNIa rate we will use the delay time distribution derived for passively evolving galaxies by \citet{Totani08}: 
$\dot N_{\rm{SN Ia}} = 0.57 (t/1\rm{Gyr})^{-1.11}$SNIa/century/$10^{10}L_{\rm{K},\odot}$ (assuming a Chabrier IMF and Z = 0.05).

The spectral shape of the ionizing background in ellipticals can then be found from the sum of two terms:

\begin{equation}
L_{\nu}(t) = L_{\nu}^{\rm{SSP}}(t) + S_{\nu}^{WD}\cdot\Delta\rm{m}_{\rm{Ia}}\cdot \dot N_{\rm{Ia}}(t)
\label{spec_L}
\end{equation}

\begin{table}
\begin{center}
\begin{tabular}{c c c c c}
\hline
\hline
{\bf Line }& $\lambda$ & {\bf SSP }& {\bf SSP+SD }& {\bf relative} \\
& & {\bf only }& {\bf channel }& {\bf enhancement }\\
\hline
\hline 
&& {\bf Ultraviolet} &&\\
\hline
$[ $C II$ ]$  & 1335.3\AA\	    & 2.0e27 & 5.6e28 & 26 \\
\hline
&& {\bf Optical} &&\\
\hline
$[ $N I$ ]$   & 5200.17\AA\	    & 4.9e27 & 2.3e29  & 45 \\
$[ $N I$ ]$   & 5197.82\AA\	    & 3.6e27 & 1.6e29 & 43 \\
$[ $O I$ ]$   & 6363.67\AA\	    & 1.3e28   & 3.0e29  & 20 \\
$[ $O I$ ]$   & 6300.2\AA\	    & 4.0e28   & 9e29  & 20 \\
\hline
&& {\bf Infra-red} &&\\
\hline
$[ $O I$ ]$   & 146$\mu$m   & 2.0e28   & 5.8e29   & 28 \\
$[ $O I$ ]$   & 63$\mu$m	    & 1.1e29   & 3.1e30  & 28 \\
$[ $C II$ ]$  & 158$\mu$m & 3.9e28    & 9.1e29  & 22 \\
\hline
\end{tabular}
\end{center}
\caption{Predicted specific line luminosities (in units of erg/s/$M_{\odot}$) for those lines most strongly effected by the introduction of a high temperature source population. We assume ionization by a 3 Gyr-old `normal' stellar population only (column 3), and with the addition of a $T_{\rm{eff}}$ = $10^{6}K$ SD SN Ia progenitor population (column 4). In both cases, we take log(U) = -3.5. \label{line_list}}
\end{table}

\noindent where $L_{\nu}^{\rm{SSP}}(t)$ is the specific ionizing luminosity of the SSP (principally arising from pAGBs), and $S_{\nu}^{WD}$ is the spectrum of radiation emitted by the population of nuclear burning white dwarfs per unit accreted mass. $S_{\nu}^{WD}$ describes the spectral energy distribution of radiation emitted by the population of SN Ia progenitors,  whose bolometric luminosity is defined by eq.(\ref{eq:ltot}). It is a superposition of spectra of individual nuclear-burning white dwarfs, each having approximately blackbody shape \citep{Rauch10} with the effective temperature determined by the mass accretion rate and the white dwarf mass. Therefore, the precise shape of $S_{\nu}^{WD}$ depends on the (poorly known) distribution of these parameters in the population of SN Ia progenitors. For the purposes of the present study, we will simply assume that $S_{\nu}^{WD}$ has a blackbody shape and can be characterized by a single photospheric temperature. In practice, any more realistic model can be constructed out of a superposition of such templates. We will then investigate how the ionization state of the ISM and the luminosity of the emission lines of interest depend on the assumed value of $T_{\rm{eff}}$. Comparison of these results with observations can be used to constrain the maximum amount of material $\Delta m_{Ia}(T_{\rm{eff}})$ which can be accreted on average by a white dwarf at various photospheric temperatures.

\begin{figure*}
\begin{center}
\hbox{
\includegraphics[height=0.25\textheight]{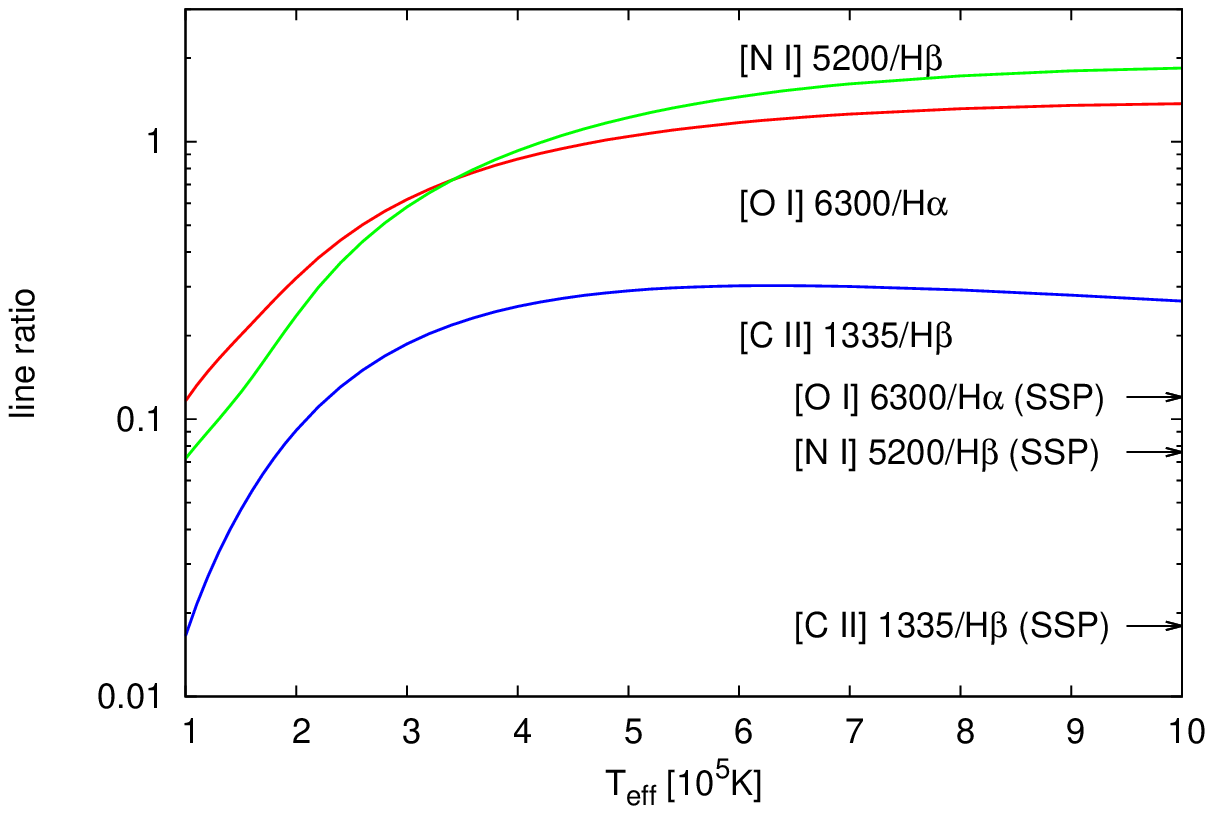}
\includegraphics[height=0.25\textheight]{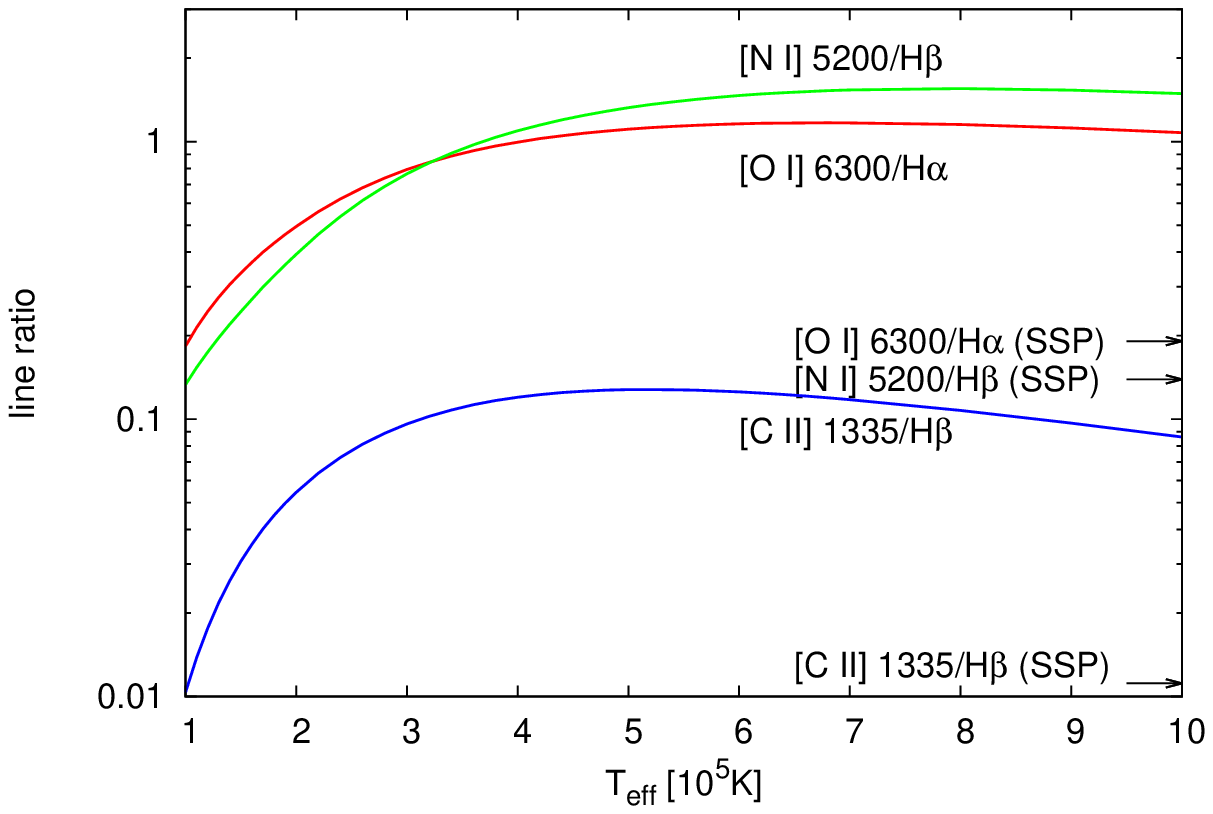}}
\caption{Predicted strength of the [C II] 1335\AA/H$\beta$, [N I] 5200\AA/H$\beta$, and [O I] 6300\AA/H${\alpha}$ ratios as a function of the effective temperature of nuclear burning WDs in the SD scenario, given ionization by both SD progenitors and pAGB stars. We assume a 3Gyr-old SSP, and a typical ionization parameter log(U) = -3.5 ({\it left}) and -4 ({\it right}), roughly spanning the expected range inferred from prior observations of early-type galaxies. Values for each ratio given ionization by the SSP alone (without the inclusion of any SD progenitor population) are labelled on each figure.}\label{OI}\label{CII}\label{NI}

\end{center}
\end{figure*}

Unless otherwise stated, we will consider in our calculations  a superposition of emission from a SSP and a high temperature SD progenitor population as a source of ionizing radiation and, when appropriate, compare obtained results with the SSP-only case. For our standard example of a high-temperature SD progenitor population, we will assume that for each SN Ia $\Delta m_{Ia}$ = 0.3$M_{\odot}$ of matter is accreted with a WD effective temperature of $T_{\rm{eff}}$ = $10^{6}K$. In those cases where we investigate the dependence of line luminosities on the WD effective temperature, we will typically consider the $10^5-10^6$ K temperature range.  As discussed in \cite{Woods13}, $\S$2.2, this covers the temperatures expected of steadily nuclear-burning WDs in the standard picture of SSSs \citep[$2\cdot 10^{5}$ -- $10^{6}K$,][]{vdHeuvel92}, as well as the so-called ``accretion-wind'' binaries with somewhat inflated photospheres \citep{HKN99}. For the present study, we do not consider any explicit dependence of either the temperatures of accreting WDs or the accreted mass $\Delta m_{\rm{Ia}}$ on the age or metallicity of the stellar population. Unless otherwise stated, we will take a 3Gyr-old SSP as representative of the relatively young passively-evolving galaxies in which we are interested, and a 500Myr SSP in our discussion of post-starburst populations.

\section{Emission-line diagnostics for high-temperature source populations}

In the standard picture of photoionized plasmas, the nebula exhibits a very sharp transition at the interface between the ionized and neutral regions \citep{Osterbrock}. 
However, this picture can be expected to change for ionization by very high temperature sources (e.g. blackbody temperatures of $\sim$ few $10^{5}K$ and higher). 
In this case, soft X-ray photons penetrate much deeper into the illuminated cloud, substantially broadening the transition region (see fig. \ref{HI_II}). 
As noted by e.g. \cite{Rappaport94}, this leads to the paradoxical conclusion that a strong enhancement in the luminosity of forbidden lines of neutral or low-ionized atoms {\bf (i.e. tracers of the Str{\" o}mgren boundary)} may be indicative of ionization by the hottest sources. 
 
The emission lines whose luminosities are most strongly enhanced (relative to the case for low-temperature ionizing sources) are those originating from neutral Nitrogen and Oxygen, and singly-ionized Carbon.  For Nitrogen and Oxygen, the first ionization energies of these relatively abundant elements ($\approx$ 14.5 eV for Nitrogen, and $\approx$ 13.6 eV for Oxygen, respectively) are very close to that of Hydrogen ($\approx$ 13.6 eV). 
Thus, emission from the forbidden transitions of [O I] and [N I] closely trace the interface of the neutral and ionized regions of nebulae.
Similarly, [C II] lines are well-known for their importance in the heating balance of largely neutral ISM \citep{Osterbrock}, given the relatively high abundance of Carbon and the low ionization potential of its valence electron.
For example, the fine structure transition $^{2}P_{1/2} \rightarrow ^{2}P_{3/2}$ is easily excited by collisions with neutral hydrogen for low gas temperatures and very low fractional ionizations \citep{Dalgarno72}.

\begin{figure*}
\begin{center}
\hbox{
\includegraphics[height=0.25\textheight]{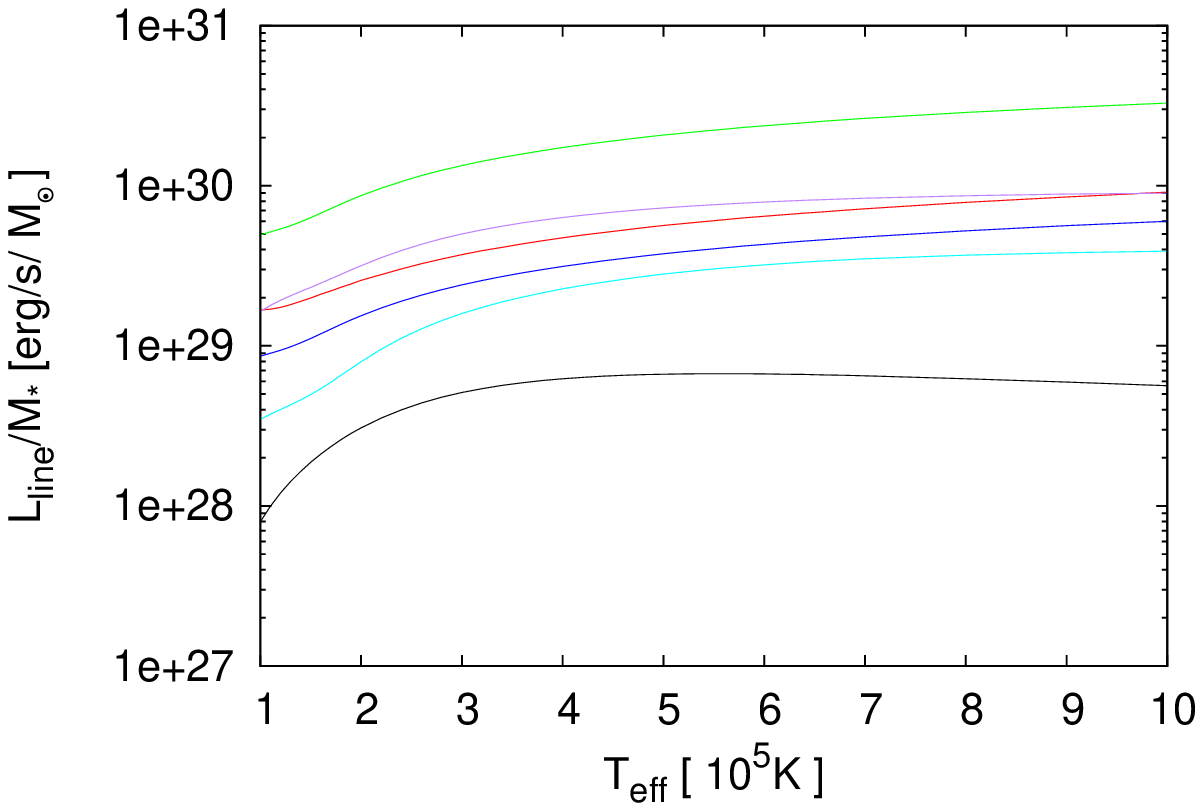}
\includegraphics[height=0.25\textheight]{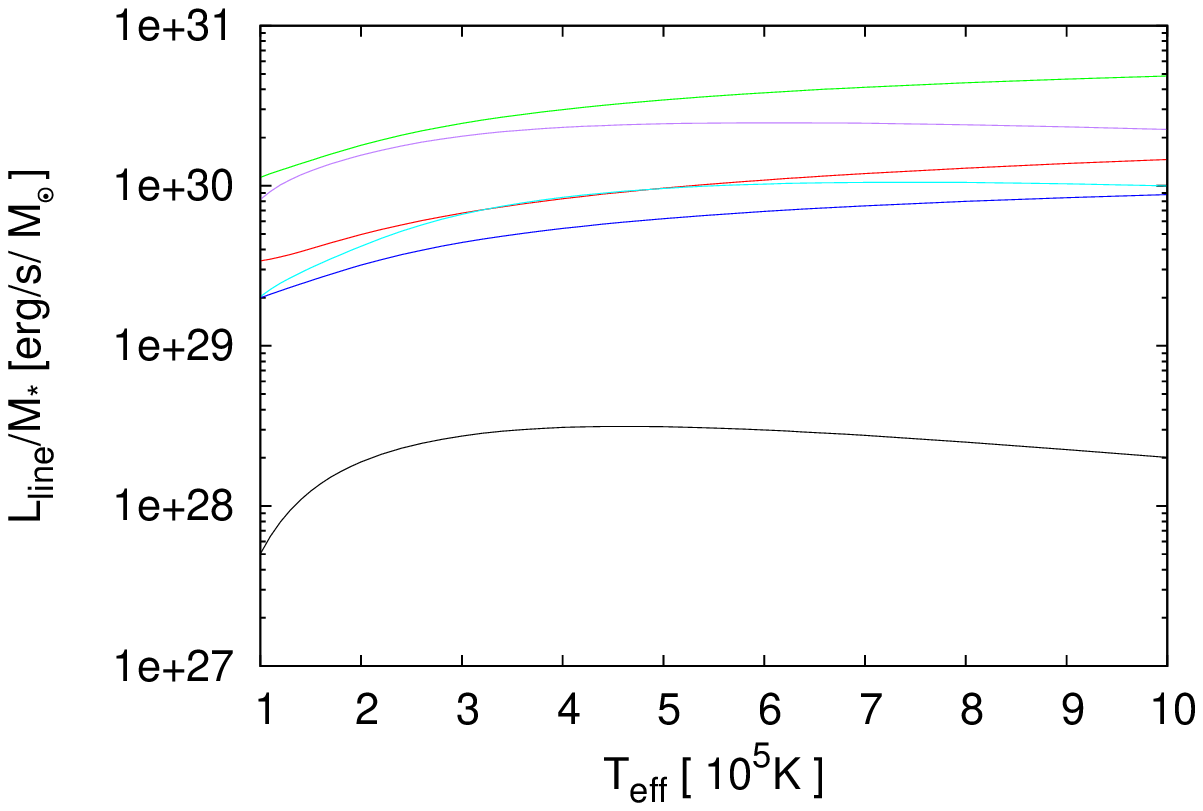}}
\caption{Predicted specific luminosities of the [C II] 157$\mu$m (red), [O I] 147$\mu$m (blue), [O I] 63$\mu$m (green), [O I] 6300\AA\ (purple), [N I] 5200\AA\ (cyan), and [C II] 1335\AA\ (black) emission lines (per unit stellar mass, for a covering fraction of unity) as a function of effective  temperature of nuclear burning WDs for our example accreting WD population, given ionization by both SD progenitors and pAGB stars. We assume an ionization parameter of log(U) = -3.5 ({\it left}) and -4 ({\it right}), for a 3Gyr-old stellar population. For clarity, we omit the line luminosities for ionization by the SSP alone from the plots. For log(U) = -3.5, and ionization by the SSP only, the line luminosities are: $L_{\rm{[C II]}1335\AA} = 2.0\cdot 10^{27} \rm{erg/s/}M_{\odot}$, $L_{\rm{[N I]}5200\AA} = 8.5\cdot 10^{27} \rm{erg/s/}M_{\odot}$, $L_{\rm{[O I]}6300\AA} = 4.0\cdot 10^{28} \rm{erg/s/}M_{\odot}$, $L_{\rm{[O I]}63\mu\rm{m}} = 1.2\cdot 10^{29} \rm{erg/s/}M_{\odot}$, $L_{\rm{[O I]}146\mu\rm{m}} = 2.0\cdot 10^{28} \rm{erg/s/}M_{\odot}$, $L_{\rm{[C II]}157\mu\rm{m}} = 3.9\cdot 10^{28} \rm{erg/s/}M_{\odot}$.}\label{abs_lums_high_T}
\end{center}
\end{figure*}

The above can easily be confirmed in a systematic manner with detailed photoionization modelling. In our calculations, we include predictions for the 50 strongest emission lines, typically extending down to line fluxes $\sim$ $10^{-6}$$L_{H\beta}$. 
As we are chiefly interested in the possible presence of very high-temperature sources within passively-evolving stellar populations, we compare ionization by the ``normally-evolving'' stellar population alone, to the same with the inclusion of our example high temperature SD progenitor population. Considering both observational errors, as well as the uncertainties inherent in the modelling of photoionized nebulae, we selected only those lines whose ratio to the nearest Hydrogen recombination line is enhanced by an order of magnitude or more.
This will allow one to meaningfully constrain the spectral hardness of the ionizing radiation in retired galaxies, allowing for the clearest distinction between ionization by pAGBs and accreting WDs.

 Selected lines are listed in table \ref{line_list}.
We have considered also intermediate temperatures of $T_{\rm{eff}}$ $\sim$ 5 $\cdot$ $10^{5}K$.
While the ordering of the lines is rearranged somewhat from table \ref{line_list}, we find that at such values for $T_{\rm{eff}}$ the principally interesting lines remain the same (with the [N I] 5200\AA\ doublet in particular remaining quite impressive). For still lower temperatures, the most interesting lines are those of He II, as discussed in \cite{Woods13}. Note that, while we assume a fixed density n = 100$\rm{cm}^{-3}$, the optical lines quoted in table \ref{line_list} vary only by $\sim$ 10 -- 20\% for densities 1$\rm{cm}^{-3}$ $<$ n $<$ 1000$\rm{cm}^{-3}$. However, the [C II] 1335\AA\ line in the UV, and particularly the IR and FIR lines of Carbon, Oxygen, and Nitrogen, are more sensitive to the density of the ionized gas (with [C II] 146$\mu$m varying by almost an order of magnitude, and the others by a factor of $\sim$ 2 over the above range in density). Therefore, our results should be used with caution outside of the density range considered here.

By far the most strongly enhanced transition is the doublet [N I] 5197.82\AA/5200.17\AA\ (hereafter we refer to their combined emission as simply [N I] 5200\AA). 
In fig. \ref{NI}, we plot the [N I] 5200\AA\ luminosity normalized to that of H$\beta$, as a function of the effective temperature for our standard case SD progenitor population. We see that for source temperatures on the order of $10^{6}K$, we expect a factor of $\approx$ 25 increase in the predicted emission-line ratio over the SSP-only case. 
Similarly, for [O I] 6300\AA/H$\alpha$, we see the predicted line ratio rise by a factor of $\approx$ 10 for the same conditions \footnote{Note that, for clarity, we omit from our plots the [O I] 6363\AA/H$\alpha$ line ratio, as it follows that of the [O I] 6300\AA\ but reduced by a factor of $\approx$ 3}.
Though less impressive, the [O I] 6300\AA\ line is intrinsically much more luminous than [N I] 5200\AA, and therefore more easily \citep[and commonly, e.g.][]{Annibali10} detected in ellipticals. 
However, the [N I] 5200\AA\ doublet is also frequently detected in the SAURON spectroscopic survey of low-ionization emission-line regions in early-type galaxies \citep[e.g.][]{SAURON06, SAURON10}. 
We reserve further discussion of observational prospects for $\S$ 5. 

In fig. \ref{CII}, we also plot the luminosity predicted in the [C II] 1335\AA\ line. Note that here we normalize our predictions to the H$\beta$ line, rather than the luminosity in the strong, nearby Ly$\alpha$ transition. This is because modeling of the total emitted luminosity of the resonant Lyman-$\alpha$ line of Hydrogen would require a careful treatment of the 3 dimensional geometry of the emitting region, in order to account for the real Ly$\alpha$ escape fraction. However, for reference, we plot an example of the Ly$\alpha$ luminosities output from MAPPINGS III in the following section (see fig. \ref{abs_lums}). Although the [C II] 1335\AA\ line is quite faint compared to the strong H and He II recombination lines in the UV, it may be an attractive target for future UV missions, such as the WSO (see $\S$ \ref{prospects}). 

As can be seen from figs. \ref{CII}, \ref{abs_lums_high_T}, above $T_{\rm{eff}}$ $\sim$ 3 -- 4$\cdot 10^{5}K$, all three of these lines are relatively insenstive to ionizing source temperature, being more or less uniformly enhanced at these temperatures. This makes them an ideal complement to the He II diagnostic \citep{Woods13}. The [N I] 5200\AA\ and [O I] 6300\AA\ lines have the added benefit of being relatively insensitive to the ionization parameter over the range relevant to retired galaxies.

We can also quantify  the effect of introducing a luminous SN Ia progenitor population in terms of the resulting absolute specific luminosities of the nebular emission lines expected in ellipticals. In fig. \ref{abs_lums_high_T}, we plot the specific luminosity in each of the optical and UV lines discussed above, as well as the three IR lines predicted to be the most enhanced in the SD scenario. 

Note that the contribution of SD progenitors to the predicted ionizing continuum is proportional to the accreted mass $\Delta m_{\rm{Ia}}$.  In principle then, an observational constraint on the luminosity of any of the above lines can be understood as an upper limit on the mass $\Delta m_{\rm{Ia}}$ which each SD progenitor accretes on average, for a given effective temperature (more on this in the $\S\S$ 4,5). 

\section{Calorimetry of stellar ionizing sources using strong emission lines}

While the most temperature-sensitive lines may provide an obvious means to test the hardness of the available ionizing spectrum, in practice the Oxygen, Nitrogen, and Carbon forbidden lines discussed above are relatively weak compared to the dominant cooling lines in photoionized nebulae. The ``classical'' strong emission lines in the optical include H$\alpha$ ($\approx$ 3H$\beta$), [N II] 6583\AA, the [O II] 3726.03\AA+3728.73\AA\ doublet (hereafter, we refer to their combined emission as [O II] 3727\AA), and [O III] 5007\AA. In the UV, the strongest line is easily the Ly$\alpha$ transition of Hydrogen. As mentioned in the previous section, the majority of these lines are relatively insensitive to the temperature of the ionizing continuum \citep{Osterbrock}, and therefore cannot constrain the precise temperature of any SD progenitor population. However their brightness makes them an easily measurable proxy for the absolute intensity of the ionizing radiation field in the galaxy, thus these lines can be used for the calorimetry of the incident flux of ionizing photons.

\begin{figure}
\begin{center}
\includegraphics[height=0.25\textheight]{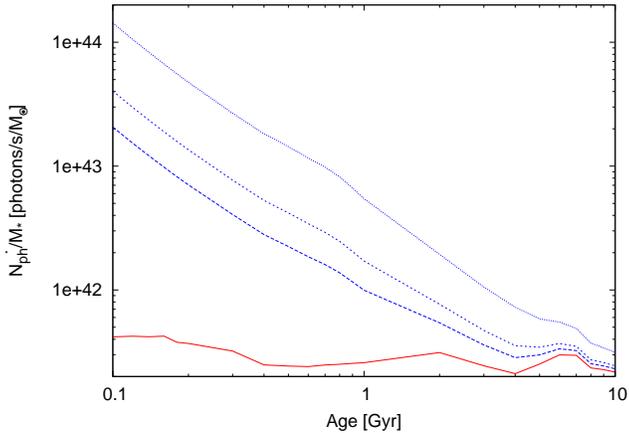}
\caption{H-ionizing photon luminosity for simple stellar populations, given ionization by pAGB stars alone (red), as well as the combined emission of both pAGB stars and SD progenitors (blue lines). Here, we assume the latter accrete $\Delta m_{\rm{Ia}}$  = $0.3M_{\odot}$ with effective temperatures of $10^{5}K$ (upper blue line), 5 $\cdot$ $10^{5}K$ (middle blue line), and $10^{6}K$ (lower blue line).}\label{0.1__1Gyr}
\end{center}
\end{figure}

Such lines are particularly useful in constraining the population of hot nuclear burning white dwarfs in very young post-starburst galaxies. This is because, for delay times less than $\sim$ 1 Gyr, we expect the total ionizing emission from SD progenitors to exceed that from the stellar population by more than an order of magnitude (see fig. \ref{0.1__1Gyr}). Therefore, the observed total emission line luminosity per unit stellar mass for {\bf any} strong line can become a useful measure of the total ionizing luminosity in young stellar populations. These lines also have the advantage of being relatively insesnsitive to the density of the ionized gas.

Given a diffuse ionizing luminosity $\rm{L}_{\nu}(t, T_{\rm{eff}}, \Delta m_{\rm{Ia}})$ produced by the stellar population (cf. eq. \ref{spec_L}), we can parametrize the luminosity in any emission line in the simple form:

\begin{equation}
\label{L_line}
\rm{L}_{\rm{line}} = \rm{f}_{\rm{c}}\cdot f_{\rm{line}}(n_{\rm{gas}}, T_{\rm{eff}}, U) \cdot \int_{13.6\rm{eV}}^{\infty} \frac{\rm{L}_{\nu}(t, T_{\rm{eff}}, \Delta m_{\rm{Ia}})}{h\nu} d\nu 
\end{equation}

\noindent where $\rm{f}_{\rm{c}}$ is the covering fraction of the ionized gas, and $T_{\rm{eff}}$ is again the effective temperature of the SD source population. Note that we scale the absolute normalization of any line to the total ionizing photon luminosity of the stellar population. The coefficient $f_{\rm{line}}$ (with dimensions of ergs per incident ionizing photon) is in principle a function of the gas density, ionizing source temperature, and ionization parameter. In this way, we encapsulate the physics relevant to any line luminosity within a simple fit to the output of our model calculations (see below). We emphasize, however, that the above parametrization is an artificial construction; for collisionally excited lines, the line luminosity is not linearly proportional to the incident ionizing photon flux, and the ionization parameter, gas density, and total ionizing photon flux are not independent quantities. Rather, this parameterization simply makes for convenient use given the input parameters of our calculations, and should not be seen as applicable outside of our treatment (ionization by blackbody sources with fixed ionization parameter and gas density).

For any recombination line, $f_{line}$ is to first approximation a trivial function of the total recombination rate and the relevant effective recombination rate for the given electronic transition. For example, the total luminosity of the H$\alpha$ line of hydrogen can be found from eq. \ref{L_line} with 

\begin{equation}
\rm{f}_{\rm{H}\alpha} = h\nu _{\rm{H}\alpha} \frac{\alpha ^{\rm{eff}}_{3\rightarrow 2}(T_{\rm{gas}})}{\alpha _{\rm{B}}} \sim 10^{-12} \rm{erg/photon}\label{Halpha}
\end{equation}

\noindent where $\alpha ^{\rm{eff}}_{3\rightarrow 2}(T_{\rm{gas}})$ is the effective rate of recombinations leading to the $3 \rightarrow 2$ transition of hydrogen, and $\alpha _{\rm{B}}$ is the total recombination rate to the 1st excited state \citep{Osterbrock}. For the approximate relation above, we assume $\rm{T}_{\rm{gas}}$ $\approx$ $10^{4}K$. Of course in practice, the gas temperature will deviate from this value. As well, the simple dependence on the relevant recombination coefficients alone breaks down for the Lyman-$\alpha$ transition of Hydrogen (Ly$\alpha$) for high-temperature ionizing sources. Here again, the difference comes about as a result of the dramatically broadened Str{\" o}mgren boundary, with ionization fraction 0 $<$ $n_{H^{+}}/n_{H}$ $<$ 1. For very hot source temperatures, high-energy photons impart significant kinetic energy to ionized electrons, which then induce secondary collisional excitations in neutral hydrogen atoms. The Ly$\alpha$ luminosity is then increasingly dominated by collisional excitations, greatly enhancing its flux over that expected from recombinations alone \citep{Shull85}.

For collisionally excited lines, the coefficient $f_{\rm{line}}$ is a non-trivial function of the gas temperature, the free electron density, and the elemental abundances. For this reason, we turn again to our detailed photoionization calculations using MAPPINGS III as above. For ionization {\bf by SD progenitors alone}, in table \ref{fits} we provide polynomial fits to the predicted line luminosities of the form:

\begin{equation}
f_{\rm{line}}(T_{\rm{eff}}) = (10^{-12} \rm{erg/photon}) \cdot \sum\limits_{n} \rm{a}_{\rm{n}}(T_{\rm{eff}}/10^{5}K)^{\rm{n}}
\label{eq:fline}
\end{equation}

\noindent where $T_{\rm{eff}}$ is the accreting WD photospheric temperature. This function is valid in the absence of any contribution from the normally-evolving stellar population, or in the limit that their contribution is negligible compared to the total luminosity of SD progenitors (which is the case, for example, for t < 1Gyr and $\Delta m_{\rm{Ia}}$ $\sim$ 0.3$M_{\odot}$). Any dependence of $L_{line}$ on the age of the stellar population will then come only from the total photon luminosity of the SD progenitor population.

It is therefore convenient to provide fits to the integrated ionizing photon luminosity in eq. \ref{L_line}, and to treat the contributions from pAGBs ($\dot N_{>13.6eV}^{SSP}$) and SD progenitors ($\dot N_{>13.6eV}^{WD}$) seperately:

\begin{equation}
\dot N_{>13.6eV}(t) =  \int\frac{L_{\nu}^{SSP}(t)}{h\nu}d\nu + \int\frac{S_{\nu}^{WD}(T_{\rm{eff}})}{h\nu}d\nu\cdot\Delta m_{\rm{Ia}}\dot N_{\rm{Ia}}(t)
\end{equation}

\noindent For nuclear-burning WDs, this can be found directly as a function of the delay time using the observed SN Ia rate, together with eq. \ref{eq:ltot} and assuming blackbody spectra. This gives us a specific ionizing photon luminosity (photons/s/$M_{\odot}$) of:

\begin{equation}
\dot N_{>13.6eV}^{WD} \approx 10^{54.2}\left(\frac{t}{\rm{yr}}\right)^{-1.37}\left(\frac{T_{\rm{eff}}}{10^{6}K}\right)^{-1}\left(\frac{\Delta M}{0.3M_{\odot}}\right) \label{SD_Ndot}
\end{equation}

\noindent for $T_{\rm{eff}}$ $\gtrsim$ 2 $\cdot$ $10^{5}K$. Note that the power-law index for the time-dependence differs slightly from the delay-time distribution; this is due to the evolution of the mass to K-band light ratio with age of the stellar population. 

The ionizing emission from the normally-evolving stellar population \citep[found from][]{BC03}, can be approximated with the polynomials:

\begin{equation}
  \dot N_{>13.6eV}^{SSP} \approx \begin{cases}
	6.5\cdot 10^{41} x^{2} - 8.6\cdot 10^{41} x + 5.2\cdot 10^{41}, \\
	 \hskip4cm\text{0.1 $\leq$ x $<$ 1.0}\\\\
	-4.3\cdot 10^{39} x + 2.8\cdot 10^{41}, \\ 
	\hskip4cm\text{1.0 $\leq$ x $\leq$ 10}.
  \end{cases}\label{pAGBs_Ndot}
\end{equation}

\noindent (for x  = (t/1 Gyr) $\lesssim$ 10) where all coefficients are in units of photons/s/$\rm{M}_{\odot}$. Note that the sum $\dot N_{>13.6eV}^{SSP} + \dot N_{>13.6eV}^{WD}$ cannot be used with the fits in table 2, as the latter were computed assuming ionization by SD progenitors alone. In practice, one would need to conduct additional calculations to include the contribution of the SSP population. However, if nececssary one can find a useful estimate for the H$\alpha$ luminosity using eq. \ref{Halpha} and eqs. \ref{SD_Ndot} and \ref{pAGBs_Ndot}.  Finally, in table \ref{fits}, we also provide power law fits for the spectral intensity of the continuum:

\begin{equation}
\rm{L}_\lambda^{\rm{cont}} = 10^{b_{0}}(t/1Gyr)^{b_{1}}\rm{erg/s/}\text{\AA}/10^{10}M_{\odot}\label{eq:continuum}
\end{equation}

\noindent This fit gives the spectral luminosity of the SSP continuum at the wavelength of each optical line, for ease in computing their equivalent widths (EW). Note that, in the case of the fit to the continuum beneath H$\alpha$, we include the Balmer absorption feature in stellar atmospheres, as given in \cite{BC03}.

Equations \ref{L_line} --  \ref{pAGBs_Ndot} provide a useful approximation for the predicted luminosity in any of the classical strong lines. For convenience, we also list in table 2 the respective coefficients for the temperature sensitive lines discussed in the previous section.

\begin{table*}
\begin{center}
\caption{Fit coefficients to the line luminosity and continuum flux at each relevant wavelength. For the former, we assume ionization by SD Progenitors alone. Valid for $10^{5}K$ $\leq$ $T_{\rm{eff}}$ $\leq$ $10^{6}K$. Values $a_{\rm{n}}$ provide the polynomial coefficients for eq. \ref{eq:fline}, which describe the respective line luminosities, while $b_{\rm{n}}$ provide the coefficients for eq. \ref{eq:continuum} at the location of the corresponding optical lines. Note that the fits given in eqs. \ref{eq:fline}, \ref{eq:continuum} are valid only for 0.1Gyr $\lesssim$ $t_{\rm{gal}}$ $\lesssim$ 1 Gyr.}
\begin{tabular}{c c c c c c c c}
\hline
Emission Line & $a_{4}$ & $a_{3}$ & $a_{2}$ & $a_{1}$ & $a_{0}$ & $b_{1}$ & $b_{0}$\\
\hline
Ly$\alpha$ 1216\AA\ & 0 & -0.13 & 2.42 & -0.37 & 7.41 & / & / \\ 
$[$O II$]$ 3727\AA\ & 0 & 0.036 & -0.72 & 5.04 & -1.16 & -1.14 & 39.67\\
$[$O III$]$ 5007\AA\ & 6.2e-2 & -0.64 & 1.91 & -0.28 & -1.7e-4 & -0.801 & 39.95\\ 
H$\alpha$ 6563\AA\ & 0 & -2.5e-3 & 5.3e-2 & -0.15  & 1.41 & -0.524 & 39.74\\ 
$[$N II$]$ 6583\AA\ & 0 & 8.6e-3 & -0.16 & 1.36 & 0.14 & -0.542 & 39.93 \\ 
\hline
$[$C II$]$ 1335\AA\  & 0 & 0 & 0 & 0.07 & -0.06 & / & / \\
$[$N I$]$  5200\AA\  & 0 & 0 & 0 & 0.33 & -0.58 & -0.720 & 40.0\\ 
$[$O I$]$  6300\AA\  & 0 & 0 & 0 & 0.85 & -1.20 & -0.574 & 39.94\\ 
$[$O I$]$  63$\mu$m  & 0 & 0 & 0 & 2.00 & -3.15 & / & / \\
$[$O I$]$  146$\mu$m & 0 & 0 & 0 & 0.37 & -0.58 & / & / \\
$[$C II$]$ 158$\mu$m & 0 & 0 & 0 & 0.41 & -0.53 & / & / \\
\hline
\end{tabular}\label{fits}
\end{center}
\end{table*}

\begin{figure*}
\begin{center}
\hbox{
\includegraphics[height=0.25\textheight]{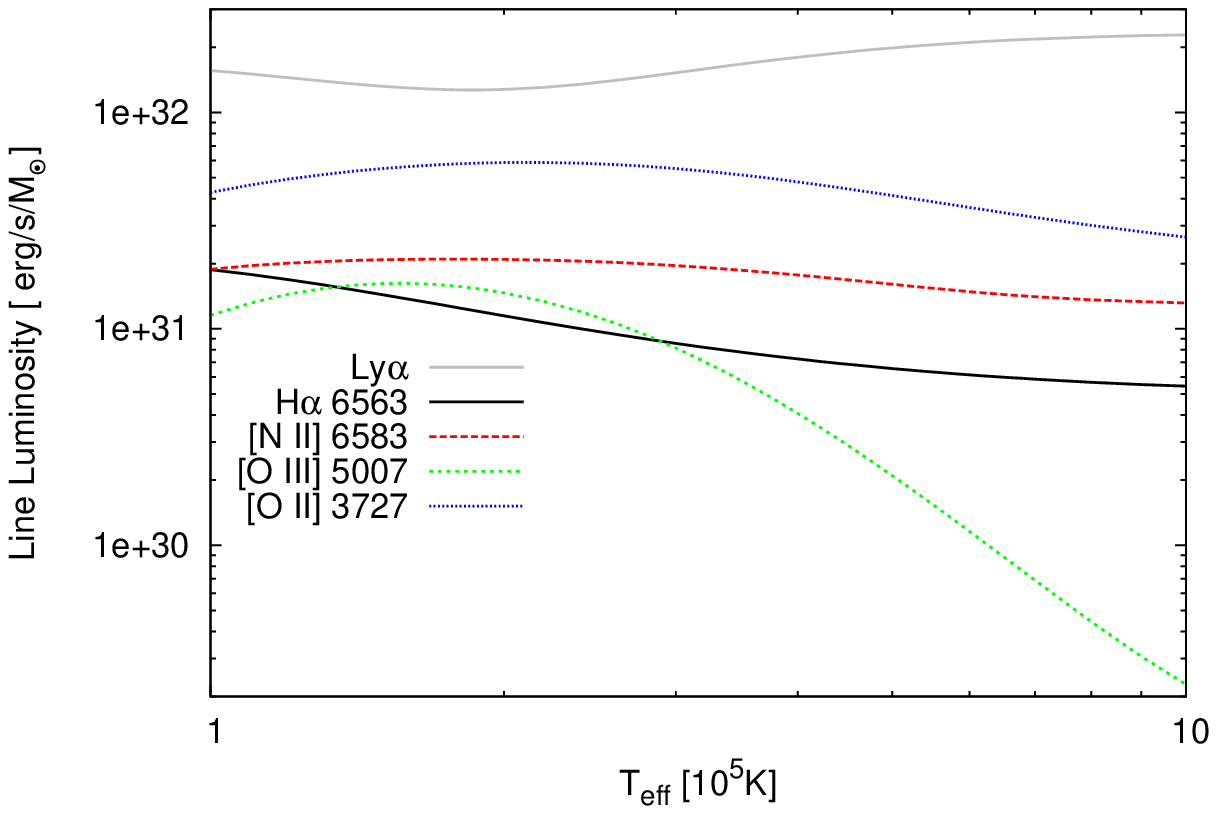}
\includegraphics[height=0.25\textheight]{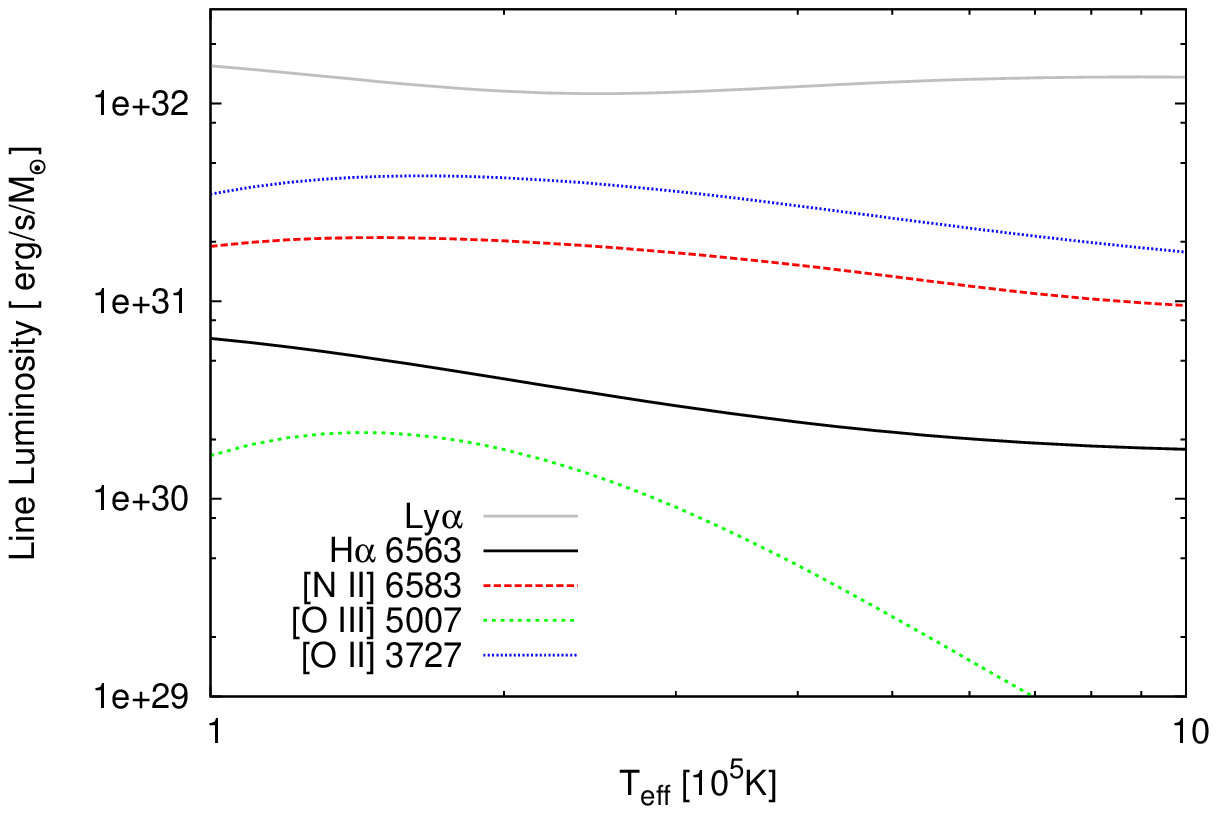}}
\caption{Absolute luminosities of the strong optical emission lines H$\alpha$, [O III] 5007\AA, [O II] 3727\AA, and [N II] 6583\AA, as a function of the photospheric temperature assumed for accreting, nuclear-burning WDs. Line fluxes are given assuming ionization by SD progenitors alone, for a stellar population of age 500Myr. We assumed $\rm{f}_{\rm{c}} = 1$ and an ionization parameter of log(U) = -3.5 ({\it left}) and -4 ({\it right}).  For clarity, we omit the line luminosities for ionization by the SSP alone from the plots. For log(U) = -3.5, the luminosities given ionization by the SSP alone are $L_{H\alpha}$ = $3.2\cdot 10^{29}$ erg/s/$M_{\odot}$, $L_{\rm{N II 6583}}$ = $3.6\cdot 10^{29}$ erg/s/$M_{\odot}$, $L_{\rm{O III 5007}}$ = $2.4\cdot 10^{29}$ erg/s/$M_{\odot}$, $L_{\rm{O II 3727}}$ = $8.5\cdot 10^{29}$ erg/s/$M_{\odot}$, and $L_{\rm{Ly}\alpha}$ = $2.7\cdot$ $10^{30}$ erg/s/$M_{\odot}$.}\label{abs_lums}
\end{center}
\end{figure*}

In figure \ref{abs_lums}, we plot the total luminosity in Ly$\alpha$, H$\alpha$, [N II] 6583\AA, [O III] 5007\AA\ and the [O II] 3727\AA\ doublet, as a function of source temperature for ionization by SD progenitors alone. Note that the predicted Ly$\alpha$ line luminosity should be only be considered an indicative estimate, as discussed in $\S$ 3. We take a 500 Myr-old SSP as our example of a post-starburst stellar population and assume our standard characteristic accreted mass $\Delta m_{\rm{Ia}} = 0.3\rm{M}_{\odot}$. From fig.5 one can  see that the presence of a high-temperature SD progenitor population leads to  an enhancement for all lines by well over an order of magnitude (except for the [O III] 5007\AA\ line for high source temperatures). The [N II] 6583\AA\  line luminosity exceeds that expected in the case of photoionization by the ``normally-evolving'' stellar population ($\approx$ 3.6 $\cdot$ $10^{29}$ $\rm{erg/s/}\rm{M}_{\odot}$ for $\rm{f}_{\rm{c}}$ = $1$) by a factor of $\approx 30-60$ across the considered temperature range. The most luminous of the considered  lines is the [O II] 3727\AA\ doublet, which lies in the near-UV on the boundary of the wavelength range typically considered the  ``optical''. However, a number of ground-based surveys have already been carried out specifically investigating this line (typically as a proxy for the SFR, see $\S$ \ref{prospects}).

\begin{figure}
\begin{center}
\includegraphics[height=0.25\textheight]{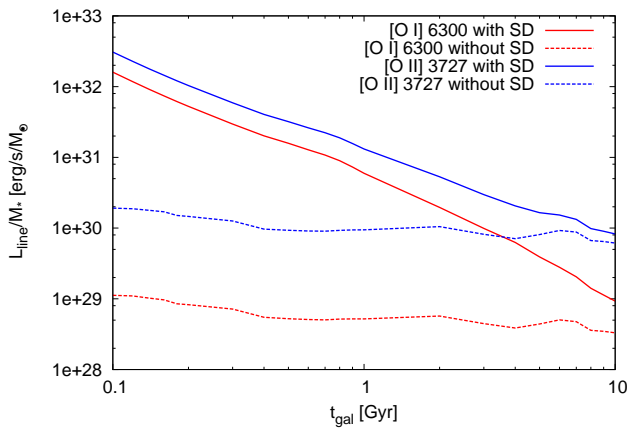}
\caption{Absolute luminosity of the [O I] 6300\AA\ line and the [O II] 3727\AA\ doublet as a function of stellar population age.  We assume $\rm{f}_{\rm{c}} = 1$ and an ionization parameter of log(U) = -3.5, with line fluxes given ionization by the ``normally-evolving'' stellar population only (red, solid), and with the inclusion of our example SD progenitor population (blue, dashed).  }\label{OII_ev_lums}
\end{center}
\end{figure}

The total luminosity in any line is determined by the incident ionizing flux, which in the SD scenario is proportional to the accreted (and nuclear-processed) mass (recall eq. \ref{eq:ltot}). Therefore any measurement (or upper limit) on their luminosities associated with a given stellar population will constrain the total ionizing luminosity of any putative SD progenitors. 
 This is particularly useful in the lower range of temperatures predicted for nuclear-burning WDs ($T_{\rm{eff}} \sim 10^{5}K$), where both the He II diagnostic and tracers of the Str{\" o}mgren boundary fail.
The significant enhancement in the line luminosity expected given photoionization by SD progenitors  should allow the total mass processed by nuclear-burning accreting WDs to be constrained to $\lesssim$ 0.01 -- 0.001 $\rm{M}_{\odot}$/SN Ia (for example, using the [O II] 3727\AA\ line, see fig. \ref{OII_ev_lums}), should line fluxes be found to be consistent with ionization by pAGBs alone.

\section{Discussion}

\subsection{Constraints on the steady nuclear-burning WD population from optical emission-line ratios}

\begin{figure*}
\begin{center}
\hbox{
\includegraphics[height=0.25\textheight]{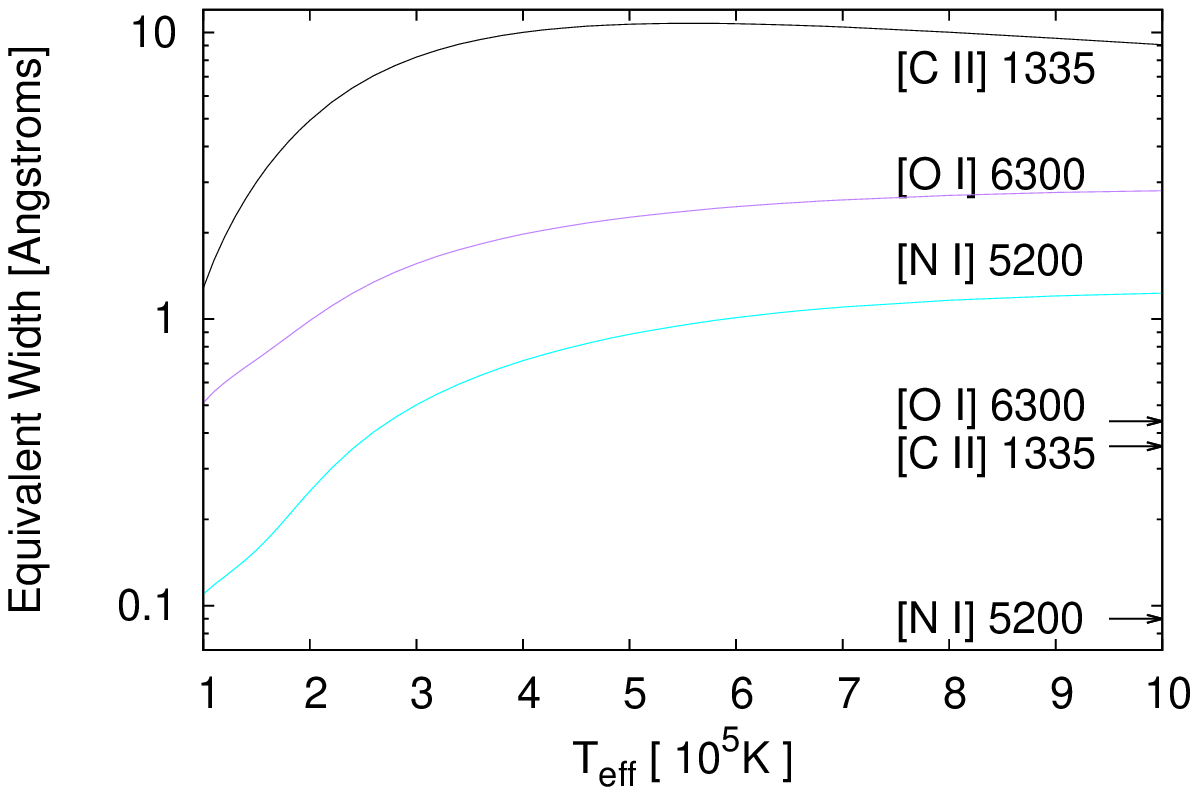}
\includegraphics[height=0.25\textheight]{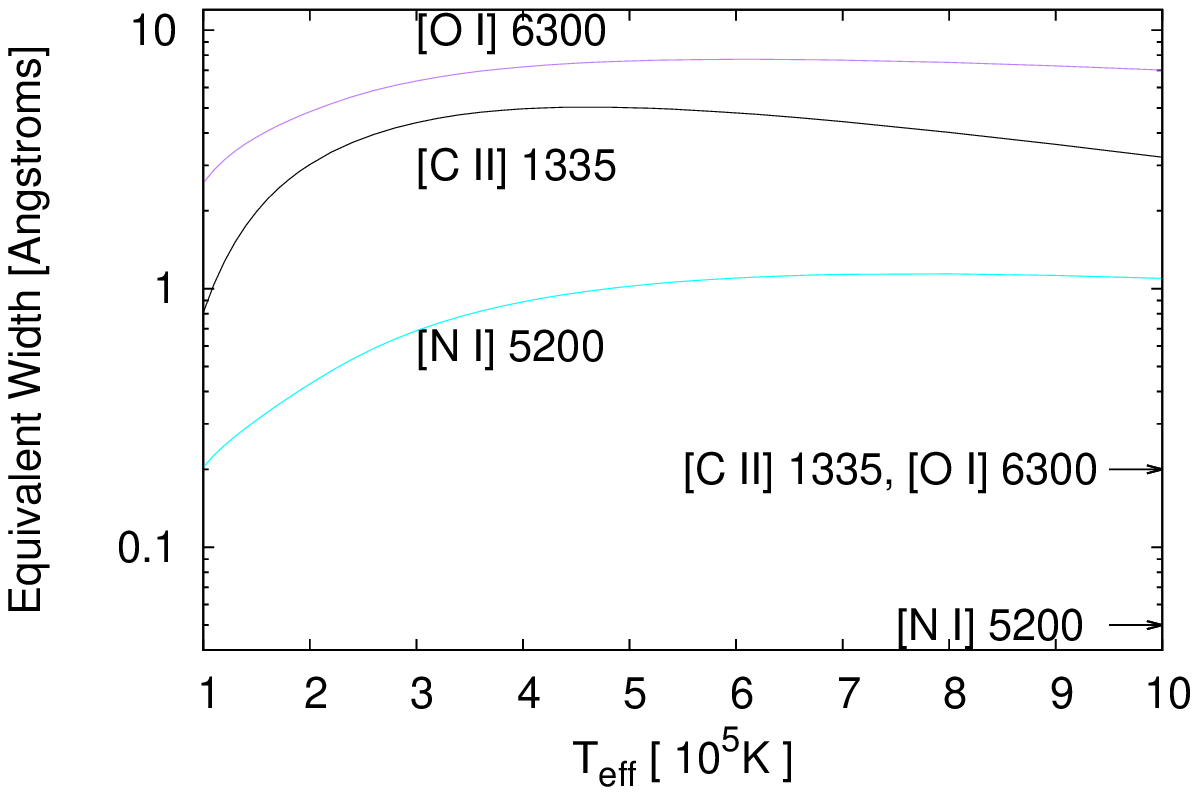}}
\caption{Equivalent widths of the [C II] 1335\AA, [N I] 5200\AA, and [O I] 6300\AA\ emission lines in the SD scenario. We assume $\rm{f}_{\rm{c}} = 1$ and an ionization parameter of log(U) = -3.5 ({\it left}) and -4 ({\it right}), for a 3-Gyr-old stellar population. For log(U) = -3.5, the EWs given ionization by the SSP alone are 0.15\AA, 0.004\AA, and 0.026\AA\ for [C II] 1335\AA, [N I] 5200\AA, and [O I] 6300\AA\ respectively. }\label{EW_highT}
\end{center}
\end{figure*}

A number of optical spectroscopic surveys of extended low-ionization emission line regions in early-type galaxies have already been conducted \citep[e.g.][]{SAURON06, Annibali10}, or are presently ongoing \citep{CALIFA}. Therefore, a large sample of such galaxies already exists which can be compared with our predictions. In estimating the feasibility of constraining the SD channel using optical forbidden lines, the most useful quantity is not the absolute luminosity per se, but rather the EW, as it characterizes the observability of the line. In fig. \ref{EW_highT}, we show for a 3 Gyr-old SSP the EW of [N I] 5200\AA, [O I] 6300\AA, and [C II] 1335\AA), taking the continuum emission as predicted from \cite{BC03}. For significantly older stellar populations ($>$ 5Gyr) our diagnostics become signigicantly less discriminating, at least for the presence of SD SN Ia progenitors, given their much lower expected luminosity at late times (recall figs. \ref{0.1__1Gyr}, \ref{OII_ev_lums}). With the EW of the line and its broadening $\sigma$, we can estimate a minimum S/N of the spectrum needed to detect any line at a desired amplitude-to-noise ratio A/N using the formula given by \cite{SAURON06}:

\begin{equation}
\frac{\rm{S}}{\rm{N}} = \frac{\rm{A/N}\hskip0.2cm\sqrt[]{2\pi}\sigma}{\rm{EW}}\label{S_N}
\end{equation}

\noindent where we will take their characteristic broadening and minimum A/N ($\sigma$ = 120km/s, A/N = 4). For SD source temperatures above $\sim 4\cdot 10^{5}K$, we predict for the [O I] line an EW of $\sim$ 1\AA\ for the realistic case $\rm{f}_{\rm{c}}$ = 0.5 (see fig. \ref{EW_highT}). Therefore, in order to barely detect [O I] 6300\AA\ emission, we would require a spectrum with S/N $\approx$ 25. Such a S/N is only just out of reach of typical SDSS spectra (S/N $\sim$ 10 -- 15), and a very robust detection should esily be possible with stacking. Note that the [O I] line is also frequently detected in individual galaxies in the sample of \cite{Annibali10}. Given the similar continuum level at 6563\AA\ and the much greater intrinsic strength of the H$\alpha$ line (see fig. \ref{abs_lums}), the latter should similarly be quite easily detectable. 

\begin{figure*}
\begin{center}
\hbox{
\includegraphics[height=0.25\textheight]{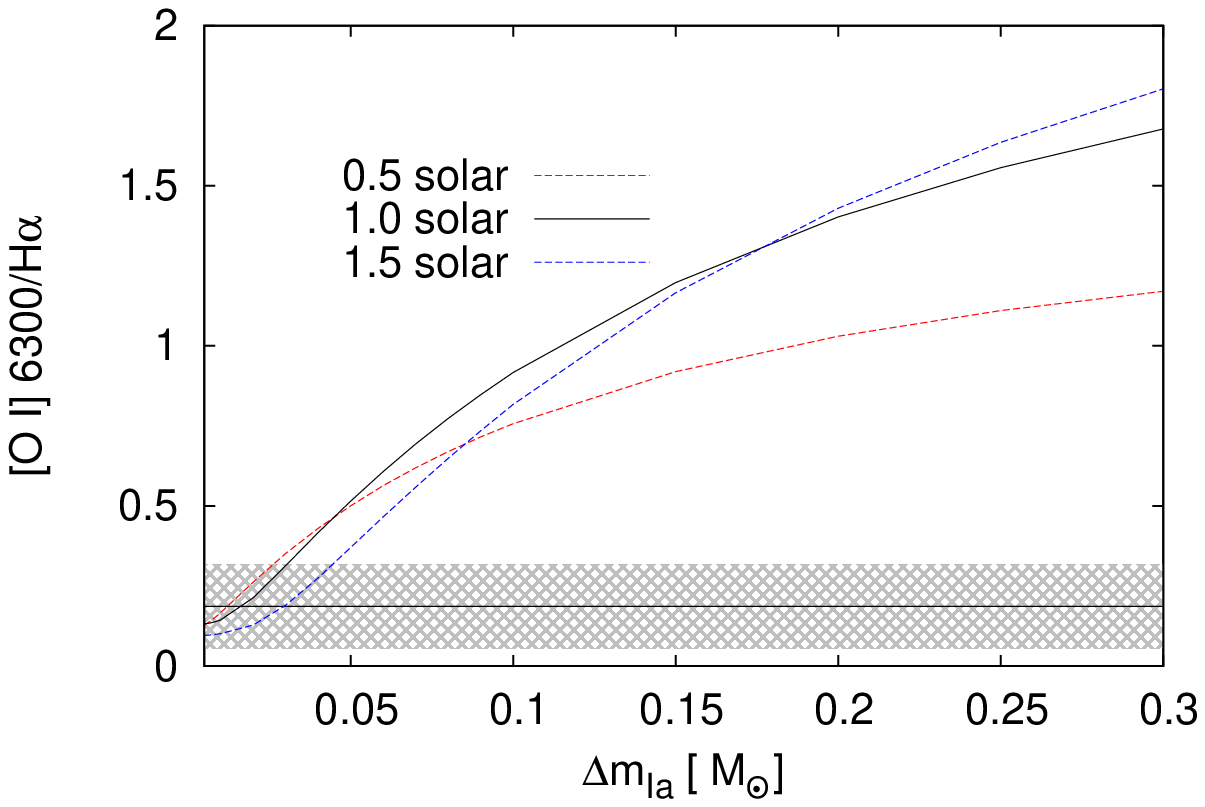}
\includegraphics[height=0.25\textheight]{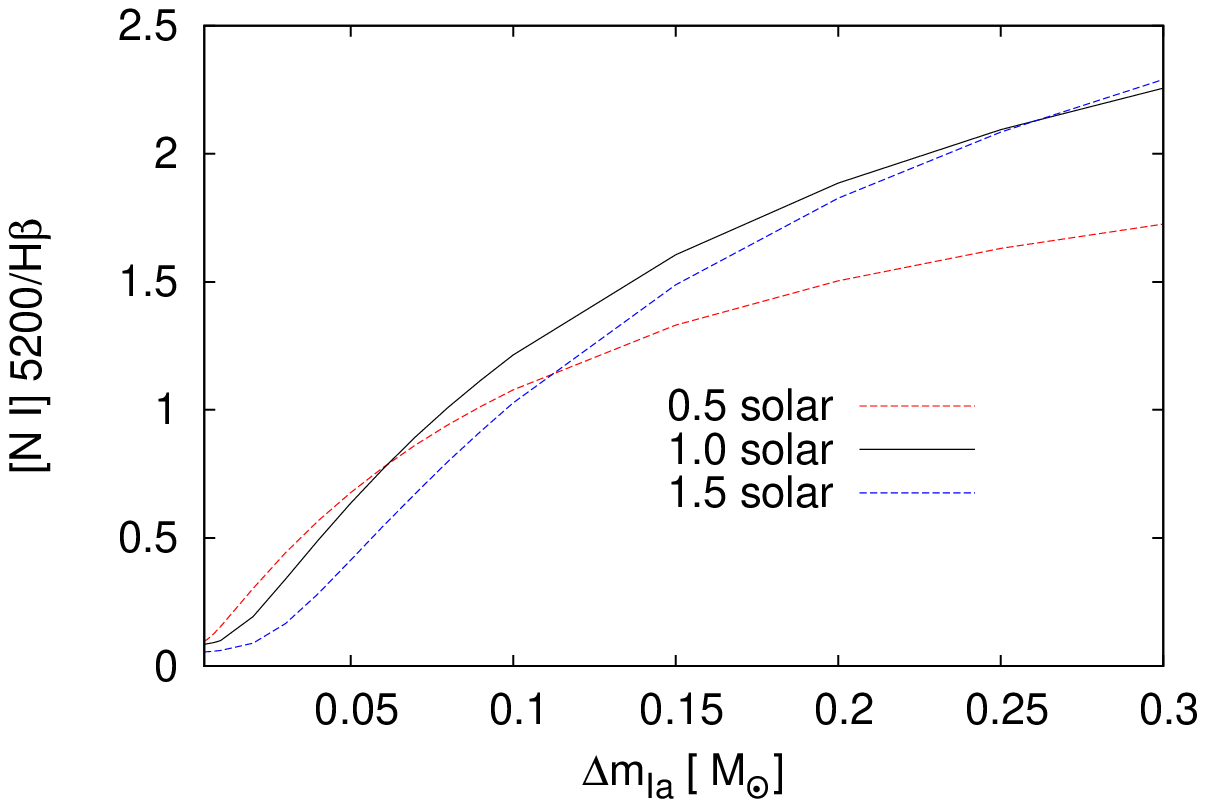}}
\caption{Predicted [O I] 6300/H${\alpha}$ ({\it left}), and [N I] 5200/H${\beta}$ ({\it right}) as a function of mass accreted by a typical WD in the SD scenario, given ionization by both SD progenitors and pAGB stars. In both figures, log(U) = -3.5, the time since initial starburst is 2 Gyr, and we assume SD progenitors with an effective temperature T = $10^{6}K$. In both figures, each of the three lines corresponds to the line luminosity assuming a different metallicity of the ionized gas, but keeping the metallicity of the stellar population fixed. For the [O I] diagram, we overplot the [O I] 6300\AA/H$\alpha$ ratio (solid black line) with 1 sigma error bars (shaded region) for the nucleus of NGC 3489 \protect\citep{Annibali10}. The [N I] 5200\AA\ doublet was not detected in the same survey.}\label{OI_dm}\label{NI_dm}
\end{center}
\end{figure*}

As discussed in $\S$ 3, normalizing our predictions for the flux in any line to that of the nearest hydrogen recombination line has the advantage of removing any dependence on the covering fraction or total mass of the stellar population. In fig. \ref{OI_dm}, we plot the predicted [O I] 6300\AA/H$\alpha$ ratio as a function of the total mass accreted by SD progenitors at an effective temperature of $10^{6}K$. Note that, whereas we would expect a linear relation for the total line flux as a function of $\Delta m_{\rm{Ia}}$, this is not the case for diagnostic line ratios. This is primarily due to the contribution of the ionizing luminosity from SD progenitors to the ionization of hydrogen. 

As an illustrative example, we can compare our models with the observed [O I] 6300\AA/H$\alpha$ ratio in the nucleus of the post-starburst galaxy NGC 3489 \citep{Annibali10, SAURON10}. As a conservative estimate \citep[see e.g.][]{Annibali07}, we assume a 2 Gyr-old SSP, with the available photoionizing continuum originating from both pAGBs and SD progenitors. We see that, for a wide range of gas phase metallicities (0.5$\rm{Z}_{\odot}$ $\leq$ $\rm{Z}$ $\leq$ 1.5$\rm{Z}_{\odot}$)\footnote{Roughly spanning the range expected in the ISM of early-type galaxies \citep{Annibali10}.}, the [O I] 6300\AA/H$\alpha$ diagnostic can in practice distinguish (at $\approx$ 1$\sigma$ confidence) between ionization by pAGBs alone and the inclusion of a SD progenitor population which accretes more than $\approx$0.04$M_{\odot}$ per SN Ia at $\rm{T}_{\rm{eff}}$ = $10^{6}K$. 

\begin{center}
\begin{figure}
\includegraphics[height=0.25\textheight]{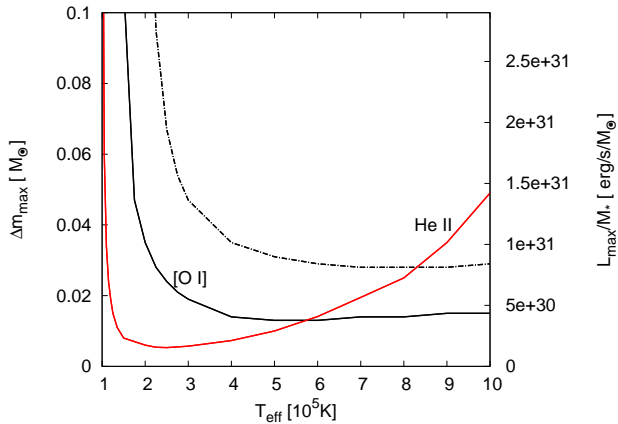}
\caption{Upper limits on the average mass $\Delta m_{\rm{Ia}}$ which may be accreted by the typical WD in the SD scenario at a given photospheric temperature. Upper limits are obtained assuming an upper limit of 10\% on the He II 4686\AA/H$\beta$ ratio (red line), and for the measured [O I] 6300\AA/H$\alpha$ ratio in the nucleus of NGC 3489 \protect\citep{Annibali10} (black, solid). For the latter, we also show 1$\sigma$ upper limit (black, dashed line). Note that any upper limit on the accreted mass is equivalent to an upper limit on the total luminosity of the population of high temperature sources (right y-axis). We assume 2 Gyr-old stellar population, log(U) = -3.5, and solar metallicity for the gas phase.}\label{dm_Teff}
\end{figure}
\end{center}

A similar case can be made for the [N I] 5200\AA/H$\beta$ line diagnostic (see fig. \ref{NI_dm}). Though the [N I] 5200\AA\ line strength is always weaker than [O I] 6300\AA\ \citep[and, unfortunately, not detected in the sample of ][]{Annibali10}, it is still observed with confidence in a number of LIERs \citep[e.g.][]{SAURON06}. In fig. \ref{EW_highT}, we see that for SD source temperatures $\gtrsim 7\cdot 10^{5}K$, the [N I] 5200\AA\ EW is roughly $\sim$ 0.5\AA\ (for $\rm{f}_{\rm{c}}=0.5$ and log(U)=-3.5). Using eq. \ref{S_N} above, we find that this line could then be detected in a 3 Gyr-old stellar population in spectra with S/N $\approx$ 40. This is well within what is typically achieved in individual SAURON spectra; with stacking, very robust detections (or upper limits) should be feasible. Stacking of SDSS spectra should allow for a confident detection with only $\sim$ 20 galaxies. From fig \ref{NI_dm}, we see that similar to the [O I] line, a detection or upper limit on the [N I] 5200\AA/H$\beta$ ratio of $\approx$ 10\% would constrain the average mass proccessed by nuclear-burning WDs in the SD channel to a few $10^{-2}M_{\odot}$ per supernova. 

Turning again to the [O I] 6300\AA/H$\alpha$ ratio in NGC 3489, we can extend the exercise in fig. \ref{NI_dm} to investigate its dependence on WD temperature. In fig. \ref{dm_Teff}, we plot the maximum mass which can be accreted by nuclear-burning WDs in the SD scenario for effective temperatures $10^{5}$ $\lesssim$ $T_{\rm{eff}}$ $\lesssim$ $10^{6}K$, should our models\footnote{Note that here, as in fig. \ref{OI_dm}, we include the expected contribution from pAGBs to the ionizing background.} remain consistent with the observed ratio [O I] 6300\AA/H$\alpha$. Note that the upper limit for any temperature in fig, \ref{dm_Teff} excludes any matter accreted at any other temperature. For example, accretion of $\sim 0.02 M_\odot$ of matter at $T_{\rm{eff}}=7\cdot 10^5$ K will produce an [O I] line luminosity equal to the observed value, therefore no more material at any other temperature could additionally be accreted without conflicting with observations. In practice, any realistic SD scenario would imply a population of WDs in varying accretion states, with a range in temperatures. Therefore any detailed SD scenario may be constrained considering the integrated luminosity and the spectrum  of the ionizing radiation produced by the population of SD progenitors.

For NGC 3489, given the observed value of the [O I] 6300\AA/H$\alpha$ ratio, no more than $\approx$ $0.015M_{\odot}$ per supernova can be accreted and nuclear-processed by WDs with photospheric temperatures $T_{\rm{eff}}$ $\gtrsim$ $4\cdot 10^{5}K$. Taking into account the [O I] line flux error, this translates to an upper limit of $\approx0.04M_{\odot}$ at $1\sigma$. Note that the flux errors in \cite{Annibali10} are not dominated by systematics, and in principle dedicated observations should be able to reduce the uncertainty considerably. If we compare this figure with the sensitivity of the He II diagnostic in \cite{Woods13} (see their fig. 8, or the red curve in our fig. \ref{dm_Teff}), we see that the two complement each other remarkably well; upper limits on the He II line luminosity (here we assume He II 4686\AA/H$\beta$ $<$ 10\%) reach their peak effectiveness at $\approx$ 2.5$\cdot 10^{5}K$, just as the [O I] 6300\AA/H$\alpha$ diagnostic becomes ineffective.  

An important point to consider is that, while we typically quote limits here in terms of the average mass accreted by nuclear-burning WDs ($\Delta m_{\rm{Ia}}$), in principle it is rather the total luminosity of any high-temperature sources which we constrain (see right y-axis in fig. \ref{dm_Teff}). Even if SD progenitors are not primarily responsible for the SN Ia rate, accreting WDs may still provide a significant contribution to the ionizing background in passively-evolving galaxies. Therefore, their consideration would be of vital importance in any exercise in understanding the role of pAGBs in low-ionization emission line regions, or their relative importance with respect to LLAGN \citep[as in, e.g. ][]{Eracleous10}. Therefore, careful modelling of the evolution of WD binaries and any phases of stable mass transfer are essential (see Chen et al, in prep).

\begin{figure}
\begin{center}
\includegraphics[height=0.25\textheight]{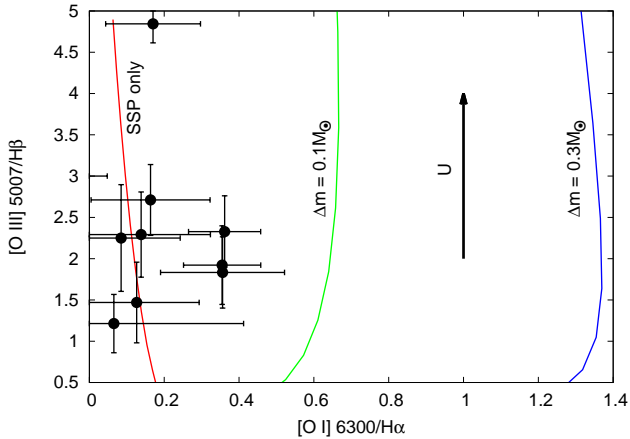}
\caption{Predicted strength of the ratios [O III] 5007\AA/H$\beta$ and [O I] 6300\AA/H$\alpha$ for varying ionization parameter, assuming ionization by a 3 Gyr-old SSP alone (red), and with the include of a SD progenitor population which accretes $\Delta m_{\rm{Ia}}$ = $0.1M_{\odot}$ (green) and $\Delta m_{\rm{Ia}}$ = $0.3M_{\odot}$ (blue) with an effective temperature $T_{\rm{eff}}$ = $10^{6}K$. The ionization parameter increases along these lines from log(U) $\approx$ -4.5 to -3, in the direction indicated by the black arrow. Shown in black are a subset of the results from \protect\cite{Annibali10}.}\label{OIII_OI}
\end{center}
\end{figure}

\begin{figure}
\begin{center}
\includegraphics[height=0.25\textheight]{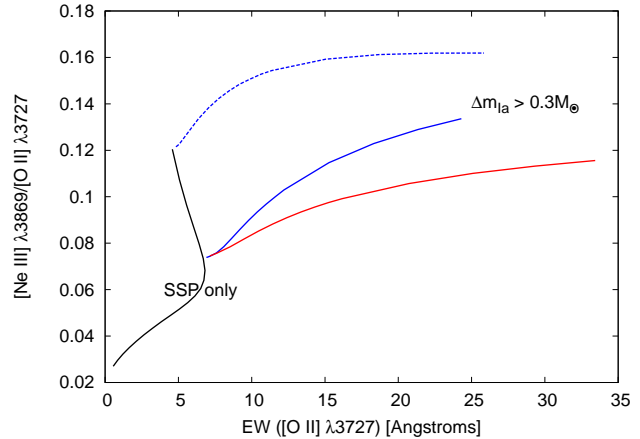}
\caption{Predicted strength of the ratio [Ne III] 3869\AA/[O II] 3727\AA and the EW of [O II] 3727\AA/H for varying ionization parameter, assuming ionization by a 3 Gyr-old SSP alone (black), and with the inclusion of a SD progenitor population which accretes varying $\Delta m_{\rm{Ia}}$ with photospheric temperature $T_{\rm{eff}}$ = $5\cdot 10^{5}K$ and log(U) = -3.5 (red, solid) and $T_{\rm{eff}}$ = $10^{6}K$ with log(U) = -3.5 (blue, solid) and log(U) = -3.0 (blue, dashed). For the tracks which include SD progenitors, $\Delta m_{\rm{Ia}}$ increases from 0 to 0.3$M_{\odot}$ from left to right in the figure.}\label{NeIII_OII}
\end{center}
\end{figure}

Finally, we note that in the presence of a hot SD progenitor population, the higher values of [O I] 6300\AA/H$\alpha$ and [N I] 5200\AA/H$\beta$ would place most passively-evolving galaxies in the region of the BPT diagram associated strictly with shock ionization (-0.5$\leq \log$([O III] 5007\AA/H$\beta$) $\leq 0.5$, and $\log$([O I] 6300\AA/H$\alpha$) $>$ -0.5). However, this has not been observed \citep[e.g.][]{SAURON10, Annibali10}. In fig. \ref{OIII_OI}, we plot the predicted values of the [O III] 5007\AA/H$\beta$ and [O I] 6300\AA/H$\alpha$ ratios with varying ionization parameter assuming ionization by a 3Gyr-old SSP alone, and with the inclusion of our standard SD progenitor population. We also include a subsample of the results from \cite{Annibali10}; namely, those galaxies with stellar population ages $<$ 4 Gyr \citep[from ][]{Annibali07}, with EW(H$\alpha$) < 1.5\AA\ \citep{Fernandes11}. This restricts us to relatively young early-type galaxies for which the ``normally-evolving'' stellar population may plausibly power the observed nebular emission. The observed sample in fig. \ref{OIII_OI} appears to be largely reproduceable with ionization by the SSP alone, although any interpretation is complicated by the considerable uncertainty in the majority of the emission line measurements.

For a number of reasons, emission-line diagnostics in the near-UV provides an attractive alternative. As discussed in the previous section, the [O II] 3727\AA\ doublet provides a much higher luminosity than H$\alpha$, in a waveband where the continuum is instrisically lower. The very nearby [Ne III] 3869\AA\ line is also available as a gauge of the ionization parameter \citep[through the \protect{[Ne III] 3869\AA/[O II] 3727\AA} ratio, e.g.][]{NeIIIOII}. This allows one to construct a BPT-style diagnostic diagram using a remarkably narrow wavelength range (as shown in fig. \ref{NeIII_OII}), using the EW of the [O II] doublet to measure the specific ionizing luminosity of the observed stellar population, and [Ne III] 3869\AA/[O II] 3727\AA\ to fix the ionization parameter. In fig, \ref{NeIII_OII}, we see that the SSP-only and SD progenitor ionizing populations occupy clearly distict regions at t = 3Gyr. For a young, post-starburst galaxy (t $\approx$ 500Myr), the EW of [O II] 3727\AA\ would be enhanced by a further order of magnitude for the same ionization parameter (see previous section). The nearby 5$\rightarrow$3 transition of He II (3203\AA)\footnote{Although $\sim$ 3 times weaker than the 4 $\rightarrow$3 transtion at 4686\AA, the continuum is predicted to be $\sim$4 times weaker at $\approx$3200\AA. Therefore the observability and utility of this transition is comparable (see eq. \ref{S_N})}. would also allow one to constrain the hardness of the ionizing source population. Together, this presents an excellent opportunity for near-UV spectroscopy to constrain the progenitors of SNe Ia.

\subsection{Constraints on the SD progenitor populations at early delay-times.}\label{prospects}

\begin{figure*}
\begin{center}
\hbox{
\includegraphics[height=0.25\textheight]{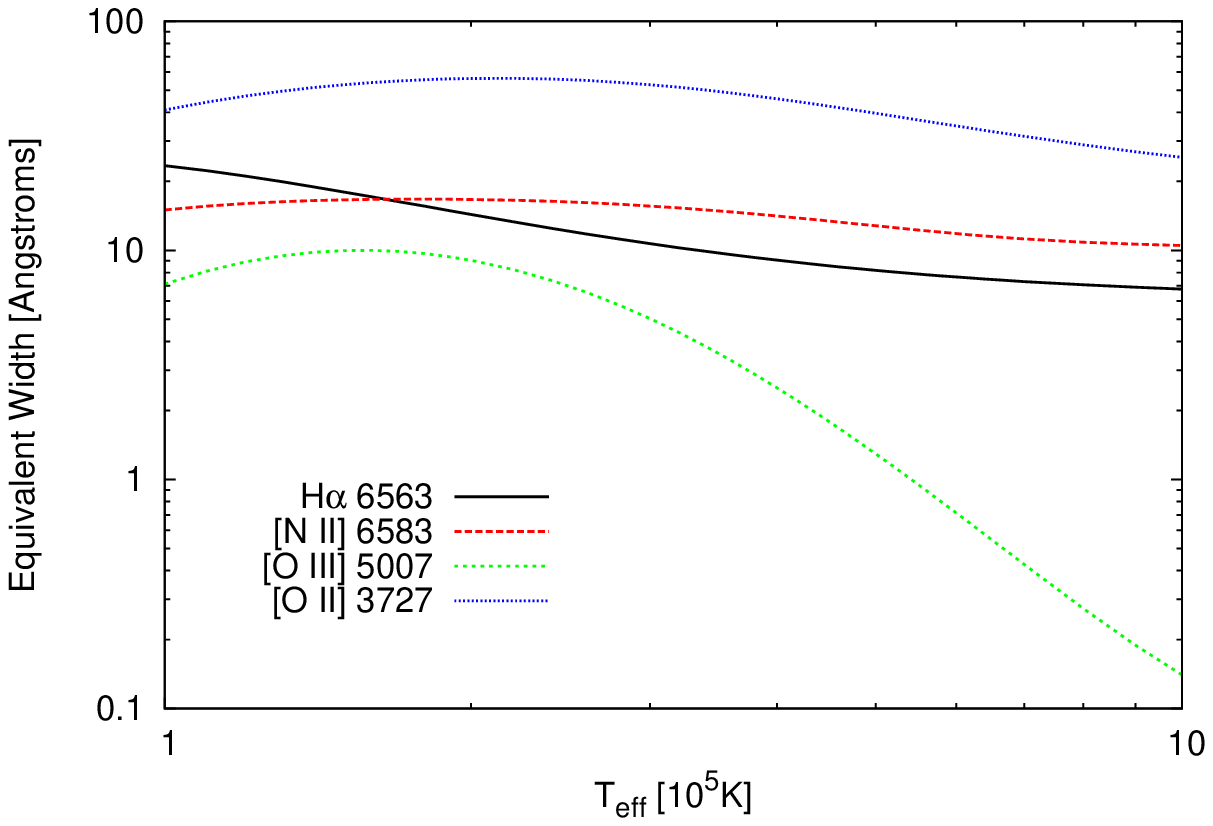}
\includegraphics[height=0.25\textheight]{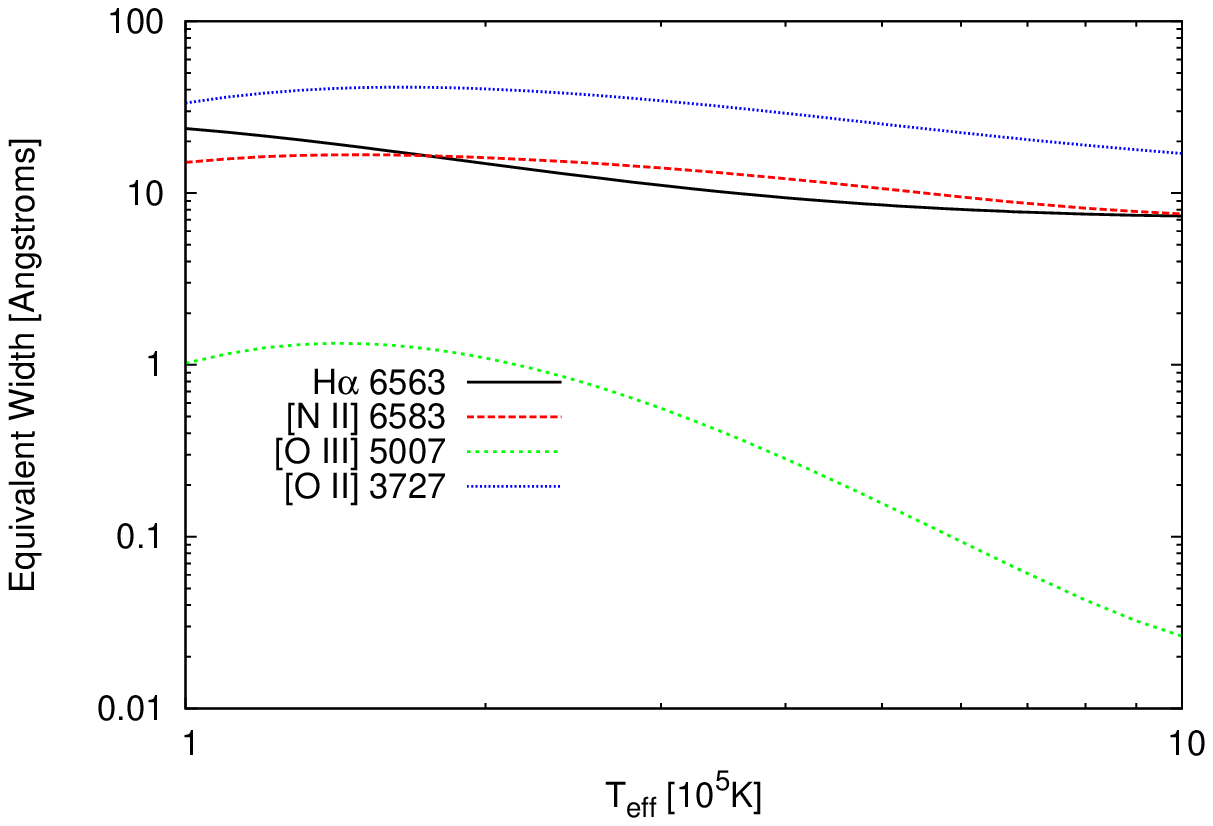}}
\caption{Equivalent widths of the strongest optical emission lines given ionization by a high temperature SD channel, assuming a 500 Myr-old stellar population. Here we assumed $\rm{f}_{\rm{c}} = 1$ and an ionization parameter of log(U) = -3.5 ({\it left}) and -4 ({\it right}). For log(U) = -3.5, the EWs given ionization by the SSP alone are 0.4\AA, 0.29\AA, 0.15\AA, and 0.8\AA\ for H$\alpha$, [N II] 6583\AA, [O III] 5007\AA, and [O II] 3727\AA\ respectively.}\label{EW_stronglines}
\end{center}
\end{figure*}

The prospect of calorimetry of the ionizing background in post-starburst galaxies offers a highly promising avenue for constraining the SD channel at the earliest delay-times. In fig. \ref{EW_stronglines}, we plot the EW of the strongest optical lines for a 500 Myr-old starburst. As discussed in $\S$ 4, the most intuitive (and least model-dependent) line for this purpose may be the H$\alpha$ recombination line. In fig. \ref{EW_stronglines}, we see that in our standard SD case, we predict an H$\alpha$ EW ranging from $\approx$ 3.5\AA\ at  $T_{\rm{eff}}=10^{6}K$ to $\approx$ 12\AA\ at $T_{\rm{eff}}=10^{5}K$, assuming $\rm{f}_{\rm{c}}$ = 0.5. This may be compared with an H$\alpha$ EW of $\approx$ 0.2\AA\ expected from the SSP alone,  giving  a factor of $\approx 20-60$  enhancement. Thus, if observations of post-starburst galaxies gave H$\alpha$ line intensities consistent with SSP-only case, the contribution of the SD channel could be strongly constrained to a few per cent level, or, equivalently, to $\Delta m_{\rm{Ia}} \lesssim 10^{-2}-10^{-3}M_\odot$ per supernova.

The [N II] 6583\AA\ line can provide a similar constraint (see fig. \ref{EW_stronglines}). However, the Nitrogen abundance in the ISM of external galaxies is known to be rather uncertain \citep{Groves04}. Therefore, there is little benefit in focussing on this line for calorimetry, unless it is unresolved from the H$\alpha$ line. The [O III] 5007\AA\ specific line luminosity drops off dramatically at the highest SD temperatures; this is because with harder spectra, Oxygen is increasingly ionized to $\rm{O}^{3+}$ near the illuminated face of the gas, reducing the emitting volume for [O III]. At the same time, high-energy photons penetrate deeper into the nebula, preferentially ionizing neutral Oxygen. Note, however, that this itself may pose a problem for any high-temperature SD channel, as the ISM in post-starburst ellipticals are observed to have relatively high values of the [O III] 5007\AA/H$\beta$ ratio \citep{SAURON10}. 

The [O II] 3727\AA\ cooling line, originating in the more low-ionized regions of nebulae (the interior, in the geometry discussed here), is expected to have an EW $\gtrsim$ 15\AA\ (for $10^{5}K$ $\lesssim$ $T_{\rm{eff}}$ $\lesssim$ $10^{6}K$). This is a $\approx 40$-fold increase over that expected from the SSP alone ($\approx 0.41$\AA, detectable with a minimum S/N $\approx$ 40). The [O II] 3727\AA\ line (as well as H$\alpha$) has been invoked as a proxy for the star-formation rate (more on this in the following section), and already the volumetric [O II] 3727\AA\ luminosity function has been mapped at least to z $\leq$ 5.4 \citep[see ][and references therein]{Gallego02}. Individual nearby galaxies which have undergone recent, possibly quenched starbursts would make ideal targets for follow-up studies. As an example, NGC 3156 is a nearby ($\approx$ 22Mpc) early-type galaxy which the SAURON survey found to host a post-starburst subpopulation (with an estimated mass of $\approx$ $10^{9}M_{\odot}$). The proposed CUBES\footnote{For the Phase A Science report, see http://www.eso.org/sci/meetings/2013/UVAstro2013/rationale.html} NUV spectrograph would be able to detect the [O II] 3727\AA\ line (given our standard SD case from above, and assuming $\rm{f}_{\rm{c}}$ = 0.5) with a S/N $\approx$ 10 $\rm{arcsec}^{-2}$ in just 1800 seconds.\footnote{Found using the CUBES Exposure Time Calculator: http://www.eso.org/observing/}

Although its specific luminosity is greatly enhanced with the presence of a high temperature source population (see fig. \ref{EW_highT}), the [C II] 1335\AA\ line remains intrinsically weak relative to the strong UV recombination lines (e.g. Ly$\alpha$, He II 1640\AA). The difficulty in detecting this line can best be illustrated with an example. At a distance of $\sim$ 21 Mpc, NGC 3607 is a nearby, relatively young \citep[mean stellar age $\sim$ 3Gyr, e.g.][]{Annibali07} elliptical galaxy which hosts an extended low-ionization emission-line region. Assuming our standard SD population from above and a covering fraction of $\rm{f}_{\rm{c}} \sim 1/2$, we can expect a total integrated [C II] 1335\AA\ line flux of $\sim$ 4.5 $\cdot$ $10^{-16}$ erg/s/$\rm{cm}^{2}$, using the K-band luminosity to estimate the mass as in \cite{Woods13}. The GALEX FUV spectrograph (now no longer operational) would have required a prohibitive $\sim$ 51 ksec in order to reach S/N $\sim$ 5,\footnote{Found using the GALEX Exposure Time Calculator (http://sherpa.caltech.edu/gips/tools/expcalc.html). In fact, this is only a minimum, as the GALEX Spectroscopy Primer recommends three times the integration time given by the ETC.} with 1335\AA\ lying outside the optimal response of the instrument. However, this remains a promising target for future UV missions, such as the planned WSO. Until such time, the aforementioned He II 1640\AA\ line remains an extremely promising target for observations \citep[see][]{Woods13}, as does the Ly$\alpha$ transition of Hydrogen at 1216\AA, the latter being easily detected in any emission-line elliptical galaxy. In the limit where SD progenitors provide the only contribution to the ionizing continuum, the total luminosity expected in Ly$\alpha$ exceeds that of the He II 1640\AA\ line by 1 -- 2 orders of magnitude for SD source temperatures $T_{\rm{eff}}$ $\approx$ $10^{5}K$ -- $10^{6}K$. 

\subsection{A Possible Bias in SFR Measures?}

In young stellar populations, the presence of strong emission lines indicates ongoing star formation \citep[in particular, for low values of N II 6583\AA/H$\alpha$, e.g.][]{Osterbrock}, where interstellar matter is ionized by highly luminous O stars. In this case, the total luminosity in any given line may be used as a tracer of the total ionizing photon flux, and therefore a measure of the number of O stars, indirectly providing the SFR (for an assumed IMF). In practice, the H$\alpha$ and [O II] 3727\AA\ fluxes are the most practical in the optical, with the standard calibrations being \citep{K98}:

\begin{equation}
\rm{SFR} (M_{\odot}/\rm{yr}) = 8 \times 10^{-42} L_{H\alpha}[erg/s]
\end{equation}

\begin{equation}
\rm{SFR} (M_{\odot}/\rm{yr}) = (1.4 \pm 0.4) \times 10^{-41} L_{[O II]}[erg/s]
\end{equation}

However, this does not account for the possibility of an additional ionizing source within the stellar population, such as possible SD progenitors of SNe Ia. This would produce an offset in the calibration of any such SFR indicator, which would be dependent on age and mass of the stellar population. For example, for $10^{5}$ $\leq$ $T_{\rm{eff}}$ $\leq$ $10^{6}K$ SD progenitor population, a 3 Gyr-old $10^{10}M_{\odot}$ SSP would still appear to have an ongoing star-formation rate of $\gtrsim \rm{f}_{\rm{c}}$ $\cdot$ 4$M_{\odot}$/yr from the [O II] 3727\AA\ indicator. Similarly, for a $T_{\rm{eff}}$ = $10^{5}K$ SD progenitor population, a 3 Gyr-old $10^{10}M_{\odot}$ SSP would still appear to have an ongoing star-formation rate of $\approx \rm{f}_{\rm{c}}$ $\cdot$ 1$M_{\odot}$/yr from the H$\alpha$ indicator.

Because of their insensitivity to interstellar reddening, IR lines and the IR continuum itself are extremely useful SFR indicators. As an important coolant in the neutral ISM, the [C II] 157$\mu$m line has been proposed as a useful SFR indicator, with the present best calibration given by \citep{CII_SFR} 

\begin{equation}
\rm{SFR} (M_{\odot}/\rm{yr}) = \frac{(L_{[C II]}[erg/s])^{0.983}}{1.028 \times 10^{40}}
\end{equation}

\noindent Yet there remains considerable scatter in the observed relation. If there is indeed a significant contribution from populations of accreting, nuclear-burning WDs, this should also provide an age-dependent offset. For example, for our standard case SD scenario, a 3 Gyr-old $10^{10}M_{\odot}$ SSP would still appear to have an ongoing star-formation rate of $\approx \rm{f}_{\rm{c}}$ $\cdot$ 1$M_{\odot}$/yr. 

The [O I] 63$\mu$m and 146$\mu$m IR lines would provide only a small contribution to any SFR-indicator based on the continuum flux. However, as individual lines, both may in principle be a useful measure of the population of accreting WDs in galaxies, in particular the 63$\mu$m line, which is $\sim$ 5 times stronger than the [C II] 157$\mu$m line for SD progenitor effective temperatures of $10^{5}$ $\leq$ $T_{\rm{eff}}$ $\leq$ $10^{6}K$. With the end of the Herschel Space Observatory's operational lifespan, there is at present no instrument which could provide new observations in this waveband, however ample archival data is available. The Field-Imaging Far-Infrared Line Spectrometer (FIFI-LS)\footnote{www.sofia.usra.edu/Science/instruments/instruments\_fifils.html}, will also soon be available, observing in the 42-210 micron band.

As a final word of caution, we recall that our results for the IR and FIR forbidden lines discussed above are sensitive to the assumed density of the gas, which is observed to vary radially \citep[e.g.][]{Yan12}. Therefore, any quantitative assessment of the contribution of a possible SD channel to the emission of these lines should consider carefully the environment under study. Furthermore, these lines do not originate solely from photoionized regions, but arise also in the neutral medium \citep{Meijerink05, Meijerink07}. 

\subsection{Mixed-age stellar populations}

So far, we have considered only single-age stellar populations, and assumed the same metallicity throughout. Although we have previously demonstrated that varying the metallicity of the stellar population will not significantly alter the ionizing continuum expected from SD progenitors \citep{Woods13}, we have not tested the effect of varying star-formation histories. In a real galaxy, a much more complex superposition of populations of differing ages should be expected. This is particularly concerning, should the presence of a significantly older subpopulation be able to strongly dilute the observed EW in a galaxy with an otherwise young mean stellar age. This could in turn significantly reduce the detectability of our diagnostic lines.

In order to evaluate this possibility, we perform the following experiment. We mix spectra of varying fractions of 1 Gyr-old, Z = 2.5$Z_{\odot}$ and 10 Gyr-old, Z = $Z_{\odot}$ stellar populations from \cite{BC03}. Using the method of \cite{Johansson12} based on the Lick indices \citep{Worthey94}, we find the luminosity-weighted age depending on the relative fractions of young and old populations (Johansson, private communication, see also Johansson et al in prep). At the same time, we compute the expected flux in any of emission lines discussed above for the composite stellar population, using the combined spectra in our photoionization calculations. Here we assume the same gas density $n = 100\rm{cm}^{-3}$ with a gas phase metallicity of Z = $\rm{Z}_{\odot}$, as above, and with ionization parameter log(U) = -3.5.

\begin{figure}
\begin{center}
\includegraphics[height=0.25\textheight]{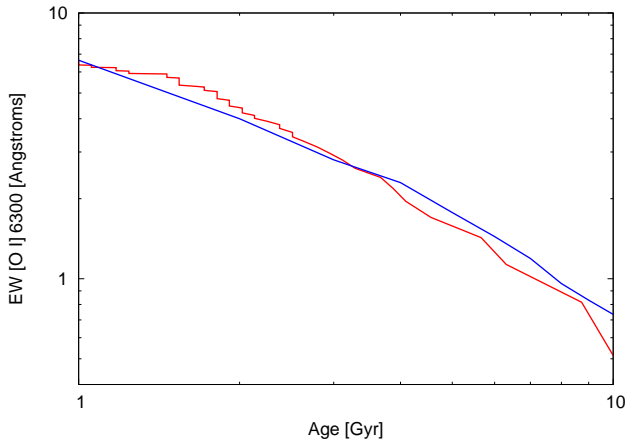}
\caption{ Equivalent width of the [O I] 6300\AA\ emission line as a function of time for single-age SSPs (blue), and for composite populations (1Gyr and 10Gyr, in red) as a function of their ''measured'' age (Johansson et al, in prep). }\label{EW_mixed_pop}
\end{center}
\end{figure}

The result of this experiment is shown in fig. \ref{EW_mixed_pop}, where we plot the EW of the [O I] 6300\AA\ emission line for such composite stellar populations, shown as a function of the derived ``mean age.'' For comparison, we also plot the [O I] EW for a single-aged SSP with Z = 0.05. Note that the two tracks do not agree at 10 Gyrs because of the different assumptions for the metallicity of the stellar population. As can be clearly seen, the EW is not significantly affected in the case of composite stellar populations.

\section{Conclusions}

We have demonstrated how observations of forbidden lines of the most abundant metals, as well as recombination lines of Hydrogen and ionized Helium, can be used to constrain any population of sources with effective temperatures $T_{\rm{eff}}$ $\gtrsim$ $10^{5}K$ in early-type galaxies. These sources may not be directly detectable due to local interstellar absorption, as well as along the line of sight in the Milky Way, but may be revealed from the dramatic effect on the emission line fluxes predicted for their host galaxies. 

A confident detection, or upper limits on, any recombination line of He II in the extended, low-ionization emission line regions of passively-evolving galaxies can constrain the total luminosity of SD progenitors with effective temperatures in the range $1.5\cdot10^{5}K$ $\lesssim$ $T_{\rm{eff}}$ $\lesssim$ $6\cdot 10^{5}K$. While this comfortably covers the range expected for symbiotics, many supersoft sourcess, and the so-called ultrasoft sources, other diagnostics must be invoked in order to constrain all possible SD progenitors. We have demonstrated that, in relatively young early-type galaxies which host significant neutral and weakly ionized  ISM, forbidden lines of [O I], [N I], and [CII] can be enhanced by up to $\sim 50$ times  in the presence of a significant ($\dot N_{\rm{Ia}}$(SD) $\approx$ $\dot N_{\rm{Ia}}$(total), very hot ($T_{\rm{eff}}$ $\gtrsim$ $5\cdot 10^{5}K$) population of accreting, nuclear-burning WDs consistent with the SD progenitor scenario. In particular, past and presently ongoing optical spectroscopic surveys should already be able to either detect evidence of the SD channel or constrain its contribution to the SN Ia rate down to a few per cent of the observed total for inferred delay-times 1 Gyr $\lesssim$ t $\lesssim$ 4 Gyr. Future observations with the planned WSO \citep[the World Space Observatory][]{WSO} should also be capable of making similar progress with observations of the [C II] 1335\AA\ (and He II 1640\AA) lines. 

The possible contribution of the SD channel at the earliest delay-times ($\lesssim 1$ Gyr) can also be remarkably well-constrained with measurements of the specific (per unit stellar mass) luminosities of the ``classical'' strong emission lines, such as H$_\alpha$, [O II] 3727 \AA, [N II] 6583\AA. For a 500 Myr-old stellar population, the H$\alpha$ recombination line luminosity is predicted to be enhanced over the SSP-only case by up to $\sim$ two orders of magnitude for SD progenitor temperatures T $\lesssim$ 2$\cdot 10^{5}K$. A similar enhancement is expected for the [O II] 3727\AA\ line luminosity for SD temperatures spanning the $10^5-10^6$ K range, suggesting calorimetry of post-starburst galaxies may be able to constrain the SD contribution to less than $\sim$ 1\%. Alternatively, observations of emission-line nebulae consistent with ionization by the old SSP alone may be interpreted as a constraint on the total mass accreted in the steady nuclear-burning regime. Assuming a hypothetical SD scenrio accounted for all SNe Ia, this would restrict the mass accreted at the $10^5-10^6$ K range of photospheric temperatures to $\lesssim$ $10^{-2}-10^{-3}M_{\odot}$/yr.

Finally, if the SD scenario did indeed account for a significant fraction of the SN Ia rate, than one might expect the resulting enhancement in the luminosities of emission-lines H$\alpha$, [O II] 3727\AA, and [C II] 157$\mu$m to introduce an age-dependent offset in their calibration as indicators of the SFR in galaxies. At the same time, one would expect nuclear-burning WDs to provide a vital ingredient in calculating the ionization budget in low-ionization emission line regions, necessitating a careful understanding of the evolution of accreting WD populations. 

\section*{Acknowledgements}

We thank the referee and Jonas Johansson for their helpful comments and discussion, and Brent Groves for making his current version of MAPPINGS III available at http://www.mpia-hd.mpg.de/$\sim$brent/mapiii.html.

\bsp

\label{lastpage}

\end{document}